\documentclass[12pt]{iopart}
\usepackage{epsfig}
\usepackage{graphicx}
\begin{document}
\newcommand{\beq}{\begin{equation}}
\newcommand{\eeq}{\end{equation}}
\newcommand{\beqa}{\begin{eqnarray}}
\newcommand{\eeqa}{\end{eqnarray}}
   \def\esim{\mathrel{\rlap{\raise2pt\hbox{$\sim$}}
    \lower1pt\hbox{$-$}}}         %equal to or approx. symbol
\def\lsim{\mathrel{\rlap{\lower4pt\hbox{\hskip1pt$\sim$}}
    \raise1pt\hbox{$<$}}}         %less than or approx. symbol
\def\gsim{\mathrel{\rlap{\lower4pt\hbox{\hskip1pt$\sim$}}
    \raise1pt\hbox{$>$}}}         %greater than or approx. symbol

\newcommand{\bsg}{${\rm b}\to {\rm s}\gamma$}

\title{Dark Matter Candidates}

\author{Lars Bergstr\"om}

\address{The Oskar Klein Centre for Cosmoparticle Physics\\ 
Department of Physics, Stockholm University\\
AlbaNova, SE-106 91 Stockholm, Sweden}
\ead{lbe@physto.se}
\begin{abstract}
An overview is given of various dark matter candidates. Among the many suggestions given in the literature, axions, inert Higgs doublet, sterile neutrinos, supersymmetric particles and
Kaluza-Klein particles are discussed. 
The situation
has recently become very interesting with new results on antimatter in the
cosmic rays having dark matter as one of the leading possible explanations.
Problems of this explanation and  possible solutions are discussed, 
and the importance 
of new measurements is emphasized. If the explanation is indeed dark matter,
a whole new field of physics, with unusual although not 
impossible mass and  interaction properties may soon open itself to 
discovery.
\end{abstract}

\maketitle

\section{Introduction: The Dark Matter Problem}
The dark matter problem has been a part of astrophysics for at least 
75 years -- since Zwicky's observation of a large velocity dispersion
 of the members of the Coma galaxy cluster \cite{zwicky}. Similarly, 
the problem of galactic rotation
curves - the stars rotate ``too fast'' to be bound by Newtonian gravity 
if all matter
is visible - can be traced back to Bacbcock's measurements of the 
Andromeda galaxy 1939 \cite{babcock}. It took, however, several 
decades before it was recognized as a real problem, and in its 
modern form it goes back to the late 1970's and early 1980's when 
the so-called cold dark matter paradigm appeared \cite{peebles} 
(in this context,
 cold means matter
moving with non-relativistic velocities when structure formed in the universe). 
Today, a wealth 
of impressive data from studies of the microwave background 
radiation, supernova distance measurements, and large-scale galaxy 
surveys have together solidified the Standard Model of cosmology, 
where structure formed through gravitational amplification of small 
density perturbations with the help of cold dark matter. Without the 
existence of dark matter 
the density contrast seen in the universe today could not have formed, 
given the small
amplitude of density fluctuations inferred from anisotropies of
 the cosmic microwave background.
  
Present-day cosmology of course also has another, mysterious component:
 a cosmological constant $\Lambda$ or a similar agent (such as time-varying 
{\em quintessence} exerting negative pressure such that the expansion of 
the universe is today accelerating). For the purpose of this article, 
however, this {\em dark energy} plays little role other than to fix the
 background metric and thus influencing late-time structure formation. 
In fact, most large-scale n-body simulations are now carried out in this 
cosmological Standard Model, the $\Lambda$CDM model. Modern models
of cosmology contain a brief period of enormously accelerated expansion, inflation, which gives a nearly scale-invariant spectrum of primordial fluctuations,
which together with the fact that the universe observationally appears to be very flat (i.e., the total energy density is equal to the critical density) are 
cornerstones of the Standard Model of cosmology.

There exist several extensive reviews of particle dark matter \cite{jkg,rpp,bhs,kam}
as well as a recent book \cite{bertone2},
in particular covering the prime candidate which has become something 
of a template for dark matter, namely the lightest supersymmetric particle.
In this review, I will focus mainly on recent developments. I will also
discuss some of the less often mentioned 
possibilities, like axion  dark matter and sterile neutrinos, and 
also some new interesting - though 
speculative, types of dark matter models that may perhaps 
explain the suprising new 
measurements of 
a large flux of positrons in the cosmic rays \cite{pamela_e,atic}. The enhanced
cross sections needed in these models, in particular the so-called Sommerfeld
enhancement, will also be discussed.

\subsection{Models for Dark Matter} 

Almost all current models of dark matter use the standard concept of quantum 
field theory to describe the properties of elementary particle candidates 
(for exceptions, see for instance \cite{unparticles,qballs}). This
means that they can be characterized by the mass and spin of the dark matter 
particle. The mass 
of proposed candidates spans a very large range, as illustrated in Table~\ref{table1}.

\begin{table}
\caption{\label{table1}Properties of various Dark Matter Candidates}
\footnotesize\rm
\begin{tabular*}{\textwidth}{@{}l*{15}{@{\extracolsep{0pt plus12pt}}l}}
\br
Type&Particle Spin&Approximate Mass Scale\\
\mr
Axion& 0&$\mu$eV-meV\\
Inert Higgs Doublet&0&50 GeV\\
Sterile Neutrino&1/2&keV\\
Neutralino&1/2&10 GeV - 10 TeV\\
Kaluza-Klein UED&1&TeV\\
\br
\end{tabular*}
\end{table}

The density of cold dark matter (CDM) is now given to an accuracy of a few percent. With $h$ being the Hubble constant today in units of 100 kms$^{-1}$Mpc$^{-1}$, the density derived fron the 5-year WMAP data \cite{wmap5} is
\beq
\Omega_{\rm CDM}h^2 = 0.1131 \pm 0.0034,  
\eeq
with the estimate  $h=0.705\pm 0.0134$.

Using the simplest type of models of thermally produced dark matter (reasonably far away from
 thresholds and branch cuts) this corresponds to an average of the annihilation
rate at the time of chemical decoupling of \cite{jkg}
\beq
\langle\sigma_A v\rangle= 2.8\cdot 10^{-26}\ {\rm cm}^3 {\rm s}^{-1}.\label{eq:simple}
\eeq
The fact that this corresponds to what one gets with a weak interaction cross section for particles of mass around the electroweak scale around a few hundrd GeV is sometimes 
coined the ``WIMP miracle'' (WIMP standing for Weakly Interacting 
Massive Particle), but it may of course be a coincidence. However, most of the
detailed models proposed for the dark matter are in fact containing WIMPs as
dark matter particles. 

The rate in Eq.~(\ref{eq:simple}) is a convenient quantity to keep in mind, 
but it has to be remarked that this is the value needed at the time of freeze-out, when the temperature was
typically of the order of $(0.05 - 0.1)M_X$ (with $M_X$ the mass of the 
dark matter particle) and the velocity $v/c\sim 0.2-0.3$. There are now 
publicly 
available computer codes \cite{ds,microm} that solve the Boltzmann equation numerically,
 taking various effects into account, such as co-annihilations 
which may change the effective 
average annihilation cross section appreciably if there are other states 
than the
one giving the dark matter particle which are nearly degenerate in
 mass. There are also  computer packages
 available (e.g., \cite{bayes})
 that can perform
joint Bayesian likelihood analysis of the probability distribution of
 combinations of parameters, in particular for supersymmetric 
dark matter models.

As we will see later, the simple Eq.~(\ref{eq:simple}) may be
modified by orders of magnitude in the halo today, for example by the Sommerfeld
enhancement -- if there are 
zero velocity bound states in the annihilating system (cold dark matter 
particles should today move with typical Galactic velocities of $v/c\sim 10^{-3}$). One should also be aware of the large astrophysical uncertainties
present when estimating the observable annihilation rate today, as it may
be influenced by the presence of substructure in the dark matter distribution,
such as discovered in large simulations of structure formation \cite{aquarius,vialactea}.

As an example, for indirect detection of gamma rays in the Galactic halo
the annihilation rate towards a direction making the angle $\Psi$ with respect
to the galactic centre is conveniently given by the factorized expression \cite{BUB}
\beq
\Phi_{\gamma}(\psi) =  0.94 \cdot 10^{-9}\left( \frac{N_{\gamma}\langle\sigma v\rangle}
{10^{-29}\ {\rm cm}^3 {\rm s}^{-1}}\right)\left( \frac{100\,\rm{GeV}}
{M_\chi}\right)^2  J\left(\psi\right)\;\rm{m}^{-2}
\;\rm{s}^{-1}\;\rm{sr}^{-1}\label{eq1}
\eeq
where the dimensionless function 
\beq
J\left(\psi\right) = \frac{1} {8.5\, \rm{kpc}} 
\cdot \left(\frac{1}{0.3\,{\rm GeV}/{\rm cm}^3}\right)^2
\int_{line\;of\;sight}\rho^2(l)\; d\,l(\psi)\;,
\label{eq:jpsi}
\eeq
with $\rho(l)$ being the dark matter density along the line of sight $l(\Psi)$.
(Note the numerical factor in Eq.~(\ref{eq1}) differs by a factor $1/2$ from that given
in the original reference \cite{BUB}; this takes into account the fact that the annihilating particles
are identical, as is the case for supersymmetric neutralinos. See the 
footnote in connection to Eq.~(21) of the publication \cite{ubel} for a detailed explanation.)

The particle physics factor $N_\gamma \langle\sigma v\rangle$, 
which is the angle and velocity averaged annihilation rate
times the number of photons created per annihilation, can usually be rather accurately
computed for a given dark matter candidate. In particular, for cross sections 
containing an $s$-wave piece, usually $\sigma v$ does not depend on velocity to
an excellent approximation, for the typical small Galactic velocities $v/c\sim 10^{-3}$
in the halo. However, the value of  $\sigma v$
may in some cases, in particular for the Sommerfeld enhanced models, depend on 
velocity or rather, after the angular average, on the velocity dispersion $v_{\rm disp}$, in the simplest case like $1/v_{\rm disp}$.
Also, $\rho^2(l)$ may vary rapidly along the line of sight if there exists
substructure in the halo dark matter distribution. Therefore, in general the
integration does not factorize as in Eq.~(\ref{eq1}) and 
has to be performed over the full phase space of the dark matter
 distribution, which for a given halo -- like that of the Milky Way -- 
unfortunately is poorly known.
  
\begin{figure}
\begin{center}
\includegraphics[width=12cm]{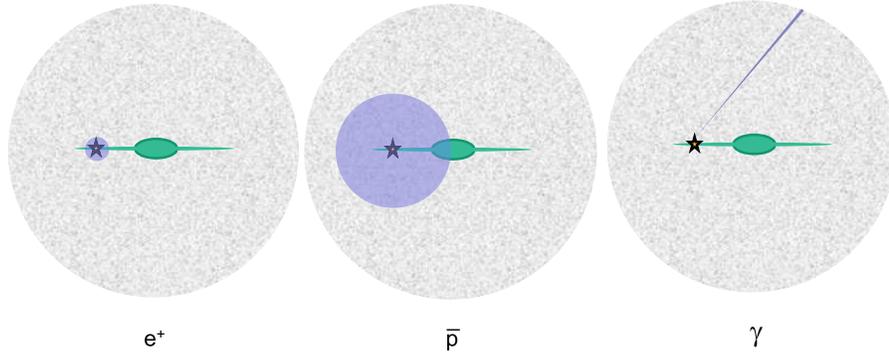}
\caption{\label{fig:fig1}Illustration of the  volumes in the solar 
neigbourhood entering the calculation of the 
average boost factor in the dark matter halo. Here we 
have in mind a dark matter particle 
of mass around 100 GeV 
annihilating
into, from left to right, positrons, antiprotons,  and gamma-rays. The 
difference in size for antiprotons and positrons depends on the different
energy loss properties, as positrons at these energies radiate through 
synchrotron and inverse Compton emission much faster than do antiprotons.}
\end{center}
\end{figure}

The procedure in this situation has been to introduce a ``boost factor'', which unfortunately does not seem to have a 
unique definition. In particular, the boost of the detection rates for a 
given model will depend on the particle of detection, and the energy. One
possible definition of the boost factor would be
\beq
B_{\rm tot}=
B_{\rho}\times B_{\sigma v}=\left({\langle \rho^2(r)\rangle_{\Delta V} \over \langle \rho_0^2(r)\rangle_{\Delta V}}\right)
\times   \left({\langle \sigma v\rangle_{v\simeq v_{\rm disp}}
 \over \langle \sigma v\rangle_{v\simeq v_F}}\right)_{\Delta_V},\label{eq:boost}
\eeq
with $v_{\rm disp}$ the velocity dispersion in the object under study and
 $v_F$ the typical velocity at freeze-out.
It is important to note that this boost factor involves an average 
of a volume $\Delta V$ which in the case of antiprotons and positrons 
would be a typical diffusion scale, and therefore could implictly 
depend on particle kind and energy (see Fig.~\ref{fig:fig1}). 

For gamma-ray observations, the enhancement 
should be computed within the line of sight cone, and therefore one 
may, for certain lines of sight, get very large boost factors, if e.g., 
these lines cross dense subhaloes (or regions, like the Galactic center, where
the influence of baryons could give an enhanced density through adiabatic contraction processes).

The computation of the boost factor in realistic astrophysical and particle 
physics scenarios is a formidable task, which has so far only been partially addressed. It may be anticipated that this will be one of the main problem areas of future indirect detection studies of dark matter. For direct detection,
there is no corresponding enhancement of the scattering rate. However, the
detailed small-scale structure of the local region of the dark matter halo may play a role \cite{Kamionkowski:2008vw}.

\subsection{Axions}\label{subs:axions}

Although at times not very much in focus of dark matter phenomenologists and experimentalists, the axion remains one of the earliest suggestions of a viable
particle candidate for dark matter, and in fact one of the most attractive.
This is not least due to the fact that its existence was motivated by solving
the strong $CP$ problem in particle physics, and its possible role for dark matter comes
as an extra bonus. A disadvantage in the cosmological context is, however, that
the axion needed to solve the $CP$ problem only solves the dark matter problem
for a small range of masses -- thus some fine-tuning mechanism seems to 
be needed. For a recent review of the field, see  \cite{axionrev}.

The original idea of Peccei and Quinn
was to make the CP violating phase dynamical \cite{PQ} by
introducing a global
symmetry, $U(1)_{PQ}$, which is
spontaneously broken. The Goldstone
boson of this broken global symmetry is the axion, which however gets a
non-zero mass from the QCD anomaly, which can be interpreted as a
mixing of the axion field with the $\pi$ and $\eta$ mesons
\cite{weinax,wilcax}.

The earliest attemps, using only the standard model particles
but with an enlarged Higgs sector (and which did not address at all the dark matter problem), were soon ruled out experimentally
and the ``invisible axion'' was invented \cite{DFSZ,KSVZ} with a very high mass
scale of symmetry breaking and with very massive fermions
carrying PQ charge. This means that  only a feeble strong or electromagnetic
interaction leaks  out to the visible sector through triangle loop
diagrams.

The
phenomenology of the axion is determined, up to numerical factors, by
one number only - the scale $f_{a}$ of symmetry breaking. In
particular, the mass is given by
\beq
m_{a}=0.62\ {\rm eV}\left({10^7\  {\rm GeV}\over f_{a}}\right).
\eeq
A naive expectation, from, e.g., Grand Unified Theories, is that $f_a$ 
is related to the  unification scale $\sim 10^{16}$ GeV, which would make the 
expected mass fall in the sub-$\mu$eV range. It turns out, however, that 
such a light axion would in general overclose the Universe and thus is not 
viable \cite{axionrev}. 
There are some hypothetical mechnisms in string theory \cite{svrcek}, 
however, that could make the mass scale smaller. It is also possible that the 
Peccei-Quinn symmetry  breaks before inflation, in which case no strong fine-tuning is required to achieve a large $f_a$. The density of axions would then 
depend on a cosmic random number (a very small misalignment angle) and anthropic selection could be unavoidable
(see, e.g., \cite{0904.0647} and references therein). A possible signature of this mechanism would be primordial isocurvature fluctuations in the cosmic microwave background \cite{0904.0647}.

The 
experimentally important coupling to two photons is due to
the effective Lagrangian term
\beq
{\cal L}_{a\gamma\gamma}
=\left({\alpha_{\rm em}\over 2\pi f_{a}}\right)\kappa\left(\vec E\cdot \vec B\right)a,\label{eq:axcoup}
\eeq
where ${\vec E}$ is the electric field, ${\vec B}$ is the magnetic field and
$\kappa$ is a model-dependent parameter of order unity.

The axion has been gradually more and more constrained by laboratory 
searches, stellar cooling
 and the supernova  dynamics 
to indeed be very light, $m_{a}< 0.01$
 eV \cite{raffelt}. It then couples so weakly to other matter
\cite{invisible} that it never was in thermal equilibrium in the early
universe and it would
 behave today as cold dark matter. There is an
 acceptable range between around $10^{-5}$ and $10^{-2}$ eV where axions
 pass all observational constraints and would not overclose the universe.

There is a considerable uncertainty in the relation
between mass and relic density,
depending on the several possible sources of axion production such as vacuum
misalignment, emission from cosmic strings etc. For a recent discussion of
the relic density of axions in various scenarios, see \cite{gondolo9}.

The coupling in Eq.~(\ref{eq:axcoup}) implies that resonant conversion
between a galactic axion
and an electric photon mode may take place in the presence of a strong
magnetic field -
not even the ``invisible axion'' may be undetectable \cite{sikivie_ax},
since the number density of these light particles in the Galaxy has to be
enormous
if axions are to make up the dark matter.

 There are now  a couple of  experiments (for a recent review, 
see \cite{bibber})
 which have had the experimental sensitivity to probe, and so 
far rule out, only a tiny part of 
the interesting region. The expected potential of the significantly upgraded Livermore experiment ADMX \cite{bibber}, will allow a deep probe into the interesting
mass window where axions are indeed a main fraction of dark matter. 

There have recently been laboratory searches \cite{cast} for light axions
emitted from processes in the Sun. Although not in a mass and coupling constant
range directly relevant for dark matter, the exclusion region covered is quite impressive
\cite{cast}. There could also be other interesting mechanisms, like 
axion-photon conversion, that could possibly influence cosmological 
measurements in interesting ways \cite{axionphoton}.

For the time being the axion remains undetected, but if it exists 
in the appropriate mass range it is still one of the prime 
candidates for the dark matter.
\subsection{Inert Higgs}

We now turn to massive particles, WIMPs, and start with one of the most minimal
extensions of the Standard Model (for an earlier, even simpler one, 
see \cite{murayama}). It was noted already in 1978 that a model with two
 Higgs doublets containing a discrete symmetry could contain a state, the lightest neutral scalar or pseudoscalar boson, which is stable \cite{ma}. Almost three decades  later, the model reappeared \cite{rychkov} as a way to obtain improved
naturalness with a Higgs that could be rather massive, larger than 300 GeV. 

The possibility of one of the lighter neutral states in the enlarged Higgs sector to be the dark matter was also pointed out, and soon the basic properties of this ``inert'' Higgs candidate for dark matter were investigated \cite{honorezlopez}. It turns out that  this model contains a dark matter candidate, a particle that does not
 couple directly to Standard Model fermions and is stable due to the discrete 
symmetry of the model (hence its relative inertness). Rates for indirect 
detection (i.e.
the observation of products of pair annihilation in the halo \cite{honorezlopez}, or in the Earth or Sun \cite{tytgat}) 
would then appear to be 
suppressed, unless its mass would be larger than the $W$ mass. However, if it is just below the $W$ mass, the virtual creation af a $W$ pair which then converts
to $\gamma\gamma$ or $Z\gamma$, would give rather spectacular rates for these
observationally interesting line processes \cite{erik1}.  These would populate
an energy region which is particularly favourable for detection in the 
Large Area Gamma-ray Telescope of the Fermi satellite \cite{fermi}. The first results on dark matter searches from Fermi should appear soon, for estimates of its potential for dark matter detection made
before launch, see \cite{prelaunch}.

The inert Higgs doublet model of dark matter is compatible with existing
accelerator bounds \cite{rychkov,erik2} and is an interesting, very minimal 
model with interesting phenomenology.

\subsection{Neutrinos}

The neutrino was a favoured particle dark matter candidate in the period 
starting in the end of the 1970's, with the first calculations of the relic density for massive neutrinos \cite{hut,leeweinberg,vysotsky,gunn}.
Of the many candidates for non-baryonic dark matter proposed,
neutrinos are often said to  have the undisputed virtue of being 
known to exist. The direct mass limits from accelerators are not very
useful for cosmological neutrinos, given the small mass differences
measured in neutrino oscillation experiments. The only direct limit 
of relevance is the one on the electron neutrino, 2 eV \cite{elnu}, which taken 
together with
mass differences inferred from neutrino observations can still allow 6 eV 
for the sum of neutrino masses. However, there are observational 
limits from cosmology on the mass range allowed for this sum, which are much more restrictive. From an analysis of the WMAP 5-year data \cite{wmap5} a bound derived on the allowed amount of a hot component translates to  
0.63 eV for the sum of neutrino masses \cite{steen}. If one is willing to also  trust
modelling of the Ly-$\alpha$ forest, a significantly better limit can
be obtained \cite{seljak}.  There has recently been progress in the rather difficult problem of treating the structure formation problem including  light neutrinos in an accurate way \cite{steenh}.

Since the contribution to the dark matter density is
\beq 
\Omega_{\nu}h^2={\sum_{i}m_{\nu_{i}}\over 94\ {\rm eV}},
\eeq 
we see that, given the cosmological bound of around 0.6 eV for the sum of neutrino masses, light neutrinos (which behave as hot, not cold, dark matter)
is at most roughly one tenth of the cold dark matter. 

There is a fundamental objection to having a massive
but light neutrino (or in fact, any fermion or boson that was once in thermal equlibrium) as
the dominant
constituent of
dark matter on all scales where it is observationally needed. This
has to do the restrictions on density evolution given by Liouville's theorem. Quantitatively, Tremaine and Gunn found \cite{tg}
that to explain the dark matter of a dwarf galaxy of velocity
dispersion $\sigma$ (usually of order 100 km/s) and core radius $r_{c}$
(typically 1 kpc), the neutrino mass has to
fulfill
\beq
m_{\nu}\geq 120\ {\rm eV}\left({100\ {\rm km/s}\over\sigma}\right)^{{1\over
4}}\left({1\ {\rm kpc}\over r_{c}}\right).\label{eq:gt}
\eeq
Recently, this bound has been improved somewhat \cite{boy}, and gives 
a lower limit of roughy 400 eV for the sterile neutrino mass. 

Sterile neutrinos do not interact
through standard weak interactions \cite{sterile}, but communicate
with the rest of the neutrino sector through fermion mixing (for recent reviews
 of
such models, see \cite{abazajian,shaposh}). 
They are limited by a
variety of observational data \cite{neulimit},
but it seems that, e.g, a region below 10 keV for
mixing angles smaller than $\sin^2\theta\sim 10^{-10}$ is allowed. 
Again one has to impose some 
unknown tuning mechanism to match the WMAP data on relic density.  
A sterile neutrino of this mass would be warm dark matter, i.e., 
intermediate between hot and cold dark matter, and this would have some
beneficial effects on some possible problems with the CDM scenario such as the 
absence of a predicted cusp in the central regions of some galaxies, or the
lack of substructure in the form of dwarf galaxies bound to the Milky Way.
Actually, the latter problem does not seem as serious now, as new data from
the Sloan Digital Sky Survey and the Keck telescope have revealed a number
of new, faint satellite galaxies \cite{simon,strigari}, and the mass dependence
of the number distribution seems to agree well with CDM simulations 
\cite{maccio}.

In principle, one could also have had a cold dark matter standard model 
neutrino 
of mass 
 around 3 GeV, but this window was closed long ago by the LEP experiment
at CERN.  
The pioneering papers which worked out the dark matter phenomenology
of such massive neutrinos (e.g., \cite{hut,leeweinberg,vysotsky,gunn})
were important, however, since they showed that a weakly interacting,
massive particle
(``WIMP'') could serve as cold dark matter with the required relic density.

To conclude this section about neutrinos,
it seems that
it is very plausible that they
make up some of the dark matter in the universe (given the
experimental results on neutrino oscillations), but most of the dark matter
is probably of some other form. Particle physics offers several other promising
candidates for this.

\subsection{Supersymmetry}

Supersymmetry has since its invention \cite{wess} fascinated a generation
of theoretical physicists, and motivated many experimentalists like those 
now involved in CERN's LHC project, likely to produce data on multi-TeV 
proton-proton collisions next year.

Supersymmetry is an ingredient in many superstring theories  which attempt to
unite all the
fundamental forces of nature, including gravity. In most versions of
the low-energy theory  there is, to avoid, for example, excessive baryon 
number violating processes, a conserved multiplicative quantum
number, R-parity:
\beq
R=\left(-1\right)^{3(B-L)+2S},
\eeq
where $B$ is the baryon number, $L$ the lepton
number and $S$ the spin of the particle. This implies that
$R=+1$ for ordinary particles
and $R=-1$ for supersymmetric particles. This means that supersymmetric
particles can only be created or annihilated in pairs in reactions
of ordinary particles. It also means that a single supersymmetric particle
can only decay into final states containing an odd number of
supersymmetric particles. In particular, this
makes the lightest supersymmetric particle
stable, since there is no kinematically allowed state with negative
R-parity which it can decay to. In fact, this is similar to the discrete
parity mentioned for the inert Higgs model. It seems that most (but not all) 
models for dark matter have to rely on a  similar discrete symmetry (which
in the simplest case is just a $Z_2$ symmetry), so maybe one should nowadays, 
when the possibility of explaining dark matter is one of the main 
motivations when 
constructing new particle physics models, generically 
introduce a multiplicative discrete ``$D$-symmetry'' (with $D$ standing for Dark) with
\beqa
D=1 & {\rm\ \ Standard\  Model\ Sector} \\
D=-1 & {\rm\ \  New\ Particle\ Sector}
\eeqa

Since this is a multiplicative quantum number, it means that particles 
in the $D=-1$ sector can only be pair-annihilated or -produced, and the lightest
particle with $D=-1$ is stable. If it is electrically neutral, it is then a dark
matter candidate.

Thus, pair-produced neutralinos $\chi$ in the early universe which
left thermal equilibrium as the universe kept expanding should have a non-zero
relic abundance today. If the scale of supersymmetry breaking is related to
that of electroweak breaking, $\chi$ will be a WIMP and 
$\Omega_{\chi}$ will be
of the right order
of magnitude to explain the non-baryonic cold dark matter.
It would indeed appear as an economic
solution if  two of the  most outstanding problems in fundamental
science, that of dark matter and that of the unification of the basic
forces, would have a common element of solution - supersymmetry.

The idea that supersymmetric particles could be good  dark matter
candidates became attractive when it was realised that breaking
of supersymmetry could be related to the electroweak scale, and
that, e.g., the supersymmetric partner of the photon (the photino)
would couple to fermions with electroweak strength \cite{fayet}.
Then
most of the phenomenology would be similar to the (failed) attemps
to have multi-GeV neutrinos as dark matter. After some early
work along these lines \cite{cabibbo,primack,weinberg2,goldberg,krauss},
the first
more complete discussion of the various possible supersymmetric
candidates was provided in \cite{ellis0}, where in particular the
lightest neutralino was identified as perhaps the most promising one.

A disadvantage of a full supersymmetric model (even making the particle content minimal, the Minimal Supersymmetric Standard Model, MSSM) is that the
number of free parameters is excessively large - of the order of 100. Therefore, most
treatments have focused on constrained models, such as minimal supergravity
(mSUGRA) models \cite{msugra},  where one has the opportunity to explain electroweak symmetry
breaking by radiative corrections caused by running from a unification scale
down to the electroweak scale (for a detailed analysis of dark matter 
in mSUGRA models, see \cite{msug_dark}).  

\subsubsection{Supersymmetric particles}\label{sec:mssm}

Let us now focus on  the lightest
supersymmetric particle, which if $R$-parity is conserved, should be stable.
In some early work, a decaying photino \cite{cabibbo}
or a gravitino \cite{primack} were considered, but for various reasons
\cite{ellis0} the most natural supersymmetric dark matter candidate  was decided to be
the lightest neutralino $\chi$. 
In fact, especially the decaying gravitino
option has recently been revived with considerable interest \cite{grav0}. 
In view of the need for very large boost factors to explain the new PAMELA \cite{pamela_e} 
and ATIC \cite{atic} data, decaying gravitino scenarios may provide
an alternative \cite{gravitino}. 

Returning to the neutralino $\chi$,
it is a mixture
of the supersymmetric partners of the photon, the $Z$ and the two neutral
$CP$-even Higgs bosons present in the minimal extension of the
supersymmetric standard model (see, e.g., \cite{haberkane}). 
It has gauge couplings
 and a mass
which for a large range of parameters in the supersymmetric sector
imply a relic density in the required range to explain
the observed $\Omega_Mh^2\sim 0.1$. As we will see, its couplings to ordinary
matter also means that its existence as dark matter in our galaxy's halo
may be experimentally tested. For an extensive review of the literature on
supersymmetric dark
matter up to mid-1995, see Ref.\,~\cite{jkg}. Some improvements were discussed
in \cite{rpp}, and the most recent, rather full discussion can be found in
\cite{bhs,bertone2}.

The phenomenology has not changed very much since these reviews.
The neutralino remains a very promising candidate, with possibilities 
for discovery in direct detection \cite{direct} and in various channels
of indirect detection \cite{indirect}.

Here we just point out two recent developments.
It was noticed in \cite{BUB} that the indirect process 
of annihilation to $\gamma\gamma$ and $Z\gamma$ 
\cite{BUB}, although often with too small branching 
ratios to be observable, 
  has a remarkable behaviour as the annihilating 
particles are either electroweak doublets (pure higgsinos) or triplets 
(pure gauginos). Namely, the cross section tends to a constant value 
proportional to $1/m_W^2$ instead as $1/m_\chi^2$ as could be expected 
on dimensional grounds. This means that the unitarity limit \cite{Jacob:1959at} 

\beq
\sigma_{\rm unitarity} < {4\pi\over vm_\chi^2}
\eeq
will eventually be violated at very high masses.
This led Hisano, Matsumoto and Nojiri \cite{Hisano:2002fk} to investigate 
the behaviour of the amplitude near that limit. They discovered that 
including perturbatively higher order corrections, they would get a 
value slightly higher than that found in \cite{BUB}, 
but more importantly, unitarity was restored. A crucial step forward 
was then taken in \cite{Hisano:2003ec} by non-perturbatively 
summing up in the ladder approximation to all orders the
 attractive $t$-channel exchange 
diagrams.  The result is a zero-energy bound state for some particular dark 
matter masses and typical galactic velocities. The appearance of the 
bound state makes the cross section increase two to three orders of 
magnitude, compared to that when velicities were corresponding to the 
freeze-out 
temperature $T\sim m_\chi/20$ (corresponding to $v\sim 0.3 - 0.4$)   
  
This phenomenon, thus discovered in \cite{Hisano:2003ec}, and verified
 in \cite{Cirelli:2005uq,Cirelli:2007xd,ArkaniHamed:2008qn,Lattanzi:2008qa}  
 (see also \cite{MarchRussell:2008yu, MarchRussell:2008tu, Hambye:2009pw}) gives the possibility of very strong indirect (in particular,
$\gamma$-ray) signals 
for particular masses (usually in the TeV region). It is 
analogous to what happens for positronium near bound state 
thresholds, as originally discussed by Sommerfeld \cite{sommerfeld}. 
It is today  the well-known ``Sommerfeld 
enhancement'' of the annihilation rate, and may be a generic phenomenon. 
Of course, supersymmetric TeV particles interacting through standard 
model gauge bosons
may have difficulty to give the required relic density, unless one
 tolerates
some fine-tuning, as is explicitly done in ``split SUSY'' models 
\cite{Giudice:2004tc}. The Sommerfeld enhancement is today 
extensively discussed in connection with the surprising new 
results on the high positron flux at high energies, see later.
If Sommerfeld enhancement is active for positrons, one would also expect 
large, perhaps detectable, signals in radio waves and
 gamma-rays \cite{Cholis:2008wq,Bertone:2008xr,bbbet}. Rather important
bounds on the enhancement follow from the early structure formation and
effects on the diffuse gamma-ray background or the cosmic microwave 
background, especially
if there is no saturation of the effect at very small velocities \cite{prokam}.

Another example of a recent development of supersymmetric dark
matter phenomenology 
(although its history goes back to the late 1980's \cite{radiative})
is the change of helicity structure caused by QED radiative corrections
to low-velocity annihilation.

This has 
recently proposed as a method to detect an otherwise undetectable
leptonic dark matter candidate \cite{Baltz:2002we}. It has also been applied to
MSSM and mSUGRA models, and found to be very important \cite{bring}, causing
sometimes large boosts to the highest energy end of the $\gamma$-ray 
spectrum. In particular, it has been shown to increase the potential for
$\gamma$-ray detection from dwarf satellite galaxies \cite{bring2}.  

A 
Majorana fermion (as many dark matter candidates are) suffers a helicity
suppression for S-wave annihilation \cite{Goldberg:1983nd}, such that the amplitude contains a factor of fermion mass $m_f$, meaning that, e.g., the $e^+e^-$ final state is highly suppressed. 
However, by emitting
a photon from an internal ($t$-channel) charged leg, which only costs a factor of
$\alpha_{em}/\pi$, the helicity suppression may be avoided. The effect will
be that these radiative corrections, instead of as usual being a percent
of the lowest order process, may instead give enhancement {\em factors}
of several thousand to million times the suppressed lowest order,
low-velocity, rate \cite{radiative}.
The resulting spectra will have a characteristic very sharp drop at
the endpoint $E_\gamma=m_\chi$ of both the $\gamma$-ray and
positron spectrum, see Fig.~2.
\begin{figure}
\begin{center}
\includegraphics[width=80mm]{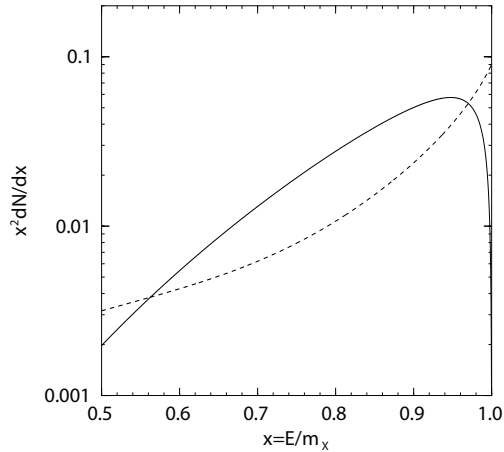}
\end{center}
\caption{\label{label}The $\gamma$-ray (solid) and positron (dashed) 
spectrum obtained by internal bremsstrahlung of the highly suppressed 
lowest order process $\chi\chi\to e^+e^-$ 
\protect\cite{radiative,bring}. Here $x=E_\gamma/m_\chi$ or $E_{e^+}/m_\chi$, and the t-channel
fermion, from which the radiation originates, is assumed to be 5 \% heavier
than the neutralino.}
\end{figure}

\subsection{Kaluza-Klein Particles in Universal Extra Dimensions}

The lightest KK particle (LKP) is an interesting, viable particle dark
matter candidate arising from extra dimensional extensions of the
standard model (SM) of particle physics. It appears in models of
universal extra dimensions (UED) \cite{appplus,matchev} 
(for the first proposal of having TeV sized extra dimensions see \cite{anton}), where all SM fields
propagate in the higher dimensional bulk, and is 
stable due to
conserved KK parity, a remnant of KK mode number conservation. This again
an example of a $D$-symmetry 
analogous to the
$R$-parity of supersymmetry, meaning the the LKP is a WIMP. Contrary to
the case of supersymmetry however, the unknown parameter space is quite
small and will be scanned entirely by, e.g., LHC.

Consider the simplest, five dimensional model with one UED
compactified on an $S^1/Z_2$ orbifold of radius $R$. All Standard Model
 fields are
then accompanied by a tower of  KK states; at
tree-level, the $n$th KK mode mass is given by
\begin{equation}
  m^{(n)} = \sqrt{(n/R)^2 + m_{\rm EW}^2}\,,
\end{equation}
where $m_{\rm EW}$ is the corresponding zero mode mass. 
The LKP can be shown in the minimal treatment of radiative corrections to be the first KK mode $B^{(1)}$ of the 
weak hypercharge gauge boson. (For a non-minimal version, see \cite{Flacke:2008ne}.)
The spectrum and the $B^{(1)}$ relic density were
first computed in \cite{matchev,ser}. Depending on the exact form of
the mass spectrum and the resulting coannihilation channels, the limit
from WMAP \cite{wmap5} of
$\Omega_{\rm CDM} h^2 = 0.1131 \pm 0.0034 $ corresponds to a mass 
of the dark matter candidate $B^{(1)}$ between roughly
0.5 to 1 TeV.  Collider measurements of electroweak observables
give a current constraint of $R^{-1} > 0.3\ {\rm TeV}$, whereas LHC should probe compactification radii up to
1.5 TeV (for a review of the detailed phenomenology of KK dark matter, 
see \cite{Hooper:2007qk}).

The most interesting aspect of KK dark matter is that it provides an example of 
a spin-1 dark matter candidate. This means that the helicity structure
of the matrix elements for annihilation will change, in particular the explicit
factor of fermion mass that appears in the $s$-wave matrix element for slow
Majorana particles annihilating in the halo is not present. This means that
new interesting direct annihilation modes will appear, such as $\nu\bar\nu$, or
$e^+e^-$, which are usually severely suppressed for neutralino annihilation, for example. The LKP was early recognized \cite{cheng,posexcess} as a potentially important source of positrons,
and was used to investigate the HEAT excess \cite{heat}, which now has been superseded
by the much more convincing PAMELA positron excess \cite{pamela_e}. In fact,
in the recent ATIC paper claiming evidence for a possible peak around 600 GeV 
perhaps related to 
dark matter \cite{atic}, a KK model was shown as an example of a model
that would give a good fit to the shape of the electron plus positron
spectrum (although there the normalization was fitted to an arbitrary, 
high value). This model has recently been revisited by 
Hooper and Zurek
\cite{Hooper:2009fj}, who conclude that a boost factor of the order of
several hundred is needed, which may be difficult to explain using current models 
of the dark matter halo. They point out, however, that such a high boost
factor need not necessarily conflict with other data at present.

\subsection{Models with Enhanced Annihilation Rate}

We have already mentioned the dramatic change of focus that has taken place
in the dark matter community, triggered by the new positron data. 
In the late summer of 2008, the first of these exciting new data were reported from 
the PAMELA
satellite, designed to measure the content of antimatter in the cosmic rays.
The results were first communicated at conferences, and later a paper was 
put on the arXiv
\cite{pamela_e}. An unexpectedly high ratio of positrons over electrons was measured, in particular in the region between 10 and 100 GeV, where previously only
weak indications of an excess had been seen \cite{oldpos}. This new precision
measurement of the cosmic ray positron flux, which definitely disagrees with 
a standard background \cite{moskstrong} has opened up a whole new field of
speculations about the possible cause of this positron excess. As mentioned, a similar, 
difficult to explain excess, a ``bump'', in the electron plus positron
spectrum was reported by the balloon experiment ATIC \cite{atic}.
Simultaneously, other data from PAMELA indicate that the antiproton
flux is in agreement with standard expectations \cite{pamela2}.

There are a variety of astrophysical models proposed for the needed extra primary component of positrons, mainly based on having nearby 
pulsars as a source \cite{pulsars}. Although pulsars with the required 
properties like distance, age, and energy output are known to exist, it turns out not to be trivial to fit both ATIC and PAMELA with these models
(see, for example, \cite{profpuls,cholpuls}). For this and other reasons,
 the dark matter interpretation,
which already had been applied to the much more uncertain HEAT data 
\cite{heatexp} has been one of the leading hyptheses (the list of relevant papers is already too long to be displayed here; 
for a partial list
of selected papers, see \cite{dm}).

It was clear from the outset that to fit the PAMELA and ATIC positron data with a dark matter  model a high mass is needed (reflecting the bump at 
around 600 GeV of ATIC). However, since the local average dark matter density
is well-known to be around 0.3 - 0.4 GeV/cm$^3$, the number density decreases
as $1/M_X$ and therefore the  annihilation rate as  
$1/M_X^2$ with $M_X$ the mass of 
the annihilating particle. This means that with $\langle\sigma v\rangle=3\cdot 10^{-26}$ cm$^3$/s, which is the standard value of the annihilation rate 
in the halo for 
thermally produced WIMPs (see Eq.~(\ref{eq:simple})), the rate of positrons,
even for a contrived model which annihilates to $e^+e^-$ with unit branching
ratio is much too small to explain the measured result.

To a good approximation, the local electron plus  positron flux for 
such a model
is  given by, assuming an energy loss of $10^{-16} E^2$ GeVs$^{-1}$ 
(with $E$ in GeV)
from inverse Compton and synchrotron radiation,
\begin{equation}
E^3{d\phi\over dE}=6\cdot 10^{-4}E
\left({1\ {\rm TeV}\over M_X}\right)^2
\theta(M_X-E)B_{\rm tot}\  
{\rm m}^{-2}
{\rm s}^{-1}{\rm sr}^{-1}{\rm GeV}^2,
\end{equation}
which means that the boost factor $B_{\rm tot}\sim 200$ (see Eq.(\ref{eq:boost}) for the definition of $B_{\rm tot}$) 
for a 600 GeV particle, 
that may otherwise explain the ATIC bump.
Similar boost factors seem to be generic, also for supersymmetric models giving
$e^+e^-$ through internal bremsstrahlung \cite{bring3}.

Returning to Eq.~(\ref{eq:boost}), we see that such a boost factor can be
given by a large inhomogeneity which has to be very local, since positrons and
electrons of several hundred GeV do not diffuse very far before 
losing essentially
all their energy, see Fig.~\ref{fig:fig1}. Although not excluded \cite{hoopa} (see, however, 
\cite{lavalle}), this would seem to be  extremely
 unlikely in most structure
formation scenarios. Therefore, most models rely on the second factor
in Eq.~(\ref{eq:boost}), i.e., the Sommerfeld enhancement factor. This means
arguably also a non-negligible amount of fine-tuning of the mass 
spectrum, in particular
also for the degeneracy between the lightest and next-to-lightest particle 
in the new sector. For a detailed discussion of the required model-builing,
see \cite{ArkaniHamed:2008qn}. Similar fine-tuning is needed for the decaying
dark matter scenario, where the decay rate has to be precisely tuned 
to give the measured flux. Since the antiproton ratio seems to be normal
according to the PAMELA measurements \cite{pamela2}, the final states should be mainly
leptons (with perhaps intermediate light new particles decaying into leptons).
For an interesting such model, which may in fact contain an almost standard
axion, see \cite{thaler}. 

It seems that at present it is possible to construct
models of the Sommerfeld enhanced type \cite{cirrai,Cholis:2008wq,Borriello:2009fa} which do marginally not 
contradict present data 
\cite{Bertone:2008xr,bbbet, Hisano:2009rc}.
We will soon, however, be presented with high 
precision data from the Fermi satellite \cite{fermi},
both for $\gamma$-rays up to 300 GeV and for the summed electron and positron spectrum
up to a TeV. Also, PAMELA and ATIC are processing further data that will 
soon be
made  public. 
It will be interesting to see whether this will give enough information  to
decide the answer to the question that at the moment is hovering in the air: {\em Has dark 
matter already been detected?}

\section*{Acknowledgements} This work was supported by the Swedish Research Council (VR). I wish to thank colleagues at the new Oskar Klein Centre in Stockholm and 
elsewhere for stimulating discussions, the Galileo Galilei Institute for Theoretical Physics in Florence for the hospitality, and the INFN for partial support during completion of this work.


\section*{References}
\begin{thebibliography}{10}
\bibitem{zwicky}
%\cite{Zwicky:1933gu}
%\bibitem{Zwicky:1933gu}
  F.~Zwicky,
  %``Spectral displacement of extra galactic nebulae,''
  Helv.\ Phys.\ Acta {\bf 6} (1933) 110.
  %%CITATION = HPACA,6,110;%%

\bibitem{babcock}
H.W. Babcock, Lick Observatory bulletin {\bf 498} (1939), 41.
\bibitem{peebles}
%\cite{Peebles:1982ff}
%\bibitem{Peebles:1982ff}
  P.~J.~E.~Peebles,
  %``Large-scale background temperature and mass fluctuations due to
  %scale-invariant primeval perturbations,''
  Astrophys.\ J.\  {\bf 263} (1982) L1.
  %%CITATION = ASJOA,263,L1;%%


\bibitem{jkg} G. Jungman, M. Kamionkowski and K. Griest, Phys. Rep.
{\bf 267} (1996) 195.
\bibitem{rpp}
%\cite{Bergstr\"om:2000pn}
%\bibitem{Bergstr\"om:2000pn}
  L.~Bergstr\"om,
  %``Non-baryonic dark matter: Observational evidence and detection methods,''
  Rept.\ Prog.\ Phys.\  {\bf 63}, 793 (2000)
  [arXiv:hep-ph/0002126].
  %%CITATION = RPPHA,63,793;%%

\bibitem{bhs} %\cite{Bertone:2004pz}
%\bibitem{Bertone:2004pz}
  G.~Bertone, D.~Hooper and J.~Silk,
  %``Particle dark matter: Evidence, candidates and constraints,''
  Phys.\ Rept.\  {\bf 405}, 279 (2005)
  [arXiv:hep-ph/0404175].
  %%CITATION = PRPLC,405,279;%%
\bibitem{kam}
%\cite{Kamionkowski:2007wv}
%\bibitem{Kamionkowski:2007wv}
  M.~Kamionkowski,
  %``Dark Matter and Dark Energy,''
  arXiv:0706.2986 [astro-ph].
  %%CITATION = ARXIV:0706.2986;%%


\bibitem{bertone2} G. Bertone ({\em ed.}), {\it Particle Dark Matter}, Cambridge University Press, 2009.
\bibitem{pamela_e}
%\cite{Adriani:2008zr}
%\bibitem{Adriani:2008zr}
  O.~Adriani {\it et al.}  [PAMELA Collaboration],
  %``Observation of an anomalous positron abundance in the cosmic radiation,''
  arXiv:0810.4995 [astro-ph].
  %%CITATION = ARXIV:0810.4995;%%

\bibitem{atic}
%\cite{:2008zzr}
%\bibitem{:2008zzr}
  J.~Chang {\it et al.},
  %``An Excess Of Cosmic Ray Electrons At Energies Of 300.800 Gev,''
  Nature {\bf 456} (2008) 362.
  %%CITATION = NATUA,456,362;%%
 
\bibitem{unparticles}
%\cite{Kikuchi:2007az}
%\bibitem{Kikuchi:2007az}
  T.~Kikuchi and N.~Okada,
  %``Unparticle Dark Matter,''
  Phys.\ Lett.\  B {\bf 665}, 186 (2008)
  [arXiv:0711.1506 [hep-ph]].
  %%CITATION = PHLTA,B665,186;%%
\bibitem{qballs}
%\cite{Kusenko:1997si}
%\bibitem{Kusenko:1997si}
  A.~Kusenko and M.~E.~Shaposhnikov,
  %``Supersymmetric Q-balls as dark matter,''
  Phys.\ Lett.\  B {\bf 418}, 46 (1998)
  [arXiv:hep-ph/9709492].
  %%CITATION = PHLTA,B418,46;%%
%\cite{Hertzberg:2008wr}
%\bibitem{Hertzberg:2008wr}
\bibitem{wmap5}
  E.~Komatsu {\it et al.}  [WMAP Collaboration],
  %``Five-Year Wilkinson Microwave Anisotropy Probe (WMAP\altaffilmark 1 )
  %Observations:Cosmological Interpretation,''
Astrophys.\ J.\ Suppl.\  {\bf 180}, 330 (2009)
  [arXiv:0803.0547 [astro-ph]].
  %%CITATION = APJSA,180,330;%%

\bibitem{ds}
%\cite{Gondolo:2004sc}
%\bibitem{Gondolo:2004sc}
  P.~Gondolo, J.~Edsj\"o, P.~Ullio, L.~Bergstr\"om, M.~Schelke and E.~A.~Baltz,
  %``DarkSUSY: Computing supersymmetric dark matter properties numerically,''
  JCAP {\bf 0407}, 008 (2004)
  [arXiv:astro-ph/0406204].
  %%CITATION = JCAPA,0407,008;%%
\bibitem{microm}
%\cite{Belanger:2004yn}
%\bibitem{Belanger:2004yn}
  G.~Belanger, F.~Boudjema, A.~Pukhov and A.~Semenov,
  %``micrOMEGAs: Version 1.3,''
  Comput.\ Phys.\ Commun.\  {\bf 174}, 577 (2006)
  [arXiv:hep-ph/0405253].
  %%CITATION = CPHCB,174,577;%%

\bibitem{bayes}
%\cite{de Austri:2006pe}
%\bibitem{de Austri:2006pe}
  R.~R.~de Austri, R.~Trotta and L.~Roszkowski,
  %``A Markov chain Monte Carlo analysis of the CMSSM,''
  JHEP {\bf 0605}, 002 (2006)
  [arXiv:hep-ph/0602028].
  %%CITATION = JHEPA,0605,002;%%

\bibitem{aquarius}
%\cite{Springel:2008zz}
%\bibitem{Springel:2008zz}
  V.~Springel {\it et al.},
  %``Prospects for detecting supersymmetric dark matter in the Galactic halo,''
  Nature {\bf 456N7218} (2008) 73.
  %%CITATION = NATUA,456N7218,73;%%

\bibitem{vialactea}
%\cite{Diemand:2006ik}
%\bibitem{Diemand:2006ik}
  J.~Diemand, M.~Kuhlen and P.~Madau,
  %``Dark matter substructure and gamma-ray annihilation in the Milky Way
  %halo,''
  Astrophys.\ J.\  {\bf 657}, 262 (2007)
  [arXiv:astro-ph/0611370].
  %%CITATION = ASJOA,657,262;%%


\bibitem{BUB}
%\cite{Bergstrom:1997fj}
%\bibitem{Bergstrom:1997fj}
  L.~Bergstr\"om, P.~Ullio and J.~H.~Buckley,
  %``Observability of gamma rays from dark matter neutralino annihilations  in
  %the Milky Way halo,''
  Astropart.\ Phys.\  {\bf 9}, 137 (1998)
  [arXiv:astro-ph/9712318].
  %%CITATION = APHYE,9,137;%%

\bibitem{ubel} 
%\cite{Ullio:2002pj}
%\bibitem{Ullio:2002pj}
  P.~Ullio, L.~Bergstr\"om, J.~Edsj\"o and C.~G.~Lacey,
  %``Cosmological dark matter annihilations into gamma-rays: A closer look,''
  Phys.\ Rev.\  D {\bf 66}, 123502 (2002)
  [arXiv:astro-ph/0207125].
  %%CITATION = PHRVA,D66,123502;%%
%\cite{Kamionkowski:2008vw}
\bibitem{Kamionkowski:2008vw}
  M.~Kamionkowski and S.~M.~Koushiappas,
  %``Galactic Substructure and Direct Detection of Dark Matter,''
  Phys.\ Rev.\  D {\bf 77}, 103509 (2008)
  [arXiv:0801.3269 [astro-ph]].
  %%CITATION = PHRVA,D77,103509;%%

\bibitem{axionrev}
  M.~P.~Hertzberg, M.~Tegmark and F.~Wilczek,
  %``Axion Cosmology and the Energy Scale of Inflation,''
  Phys.\ Rev.\  D {\bf 78}, 083507 (2008)
  [arXiv:0807.1726 [astro-ph]].
  %%CITATION = PHRVA,D78,083507;%%
\bibitem{PQ} R. Peccei and H.R. Quinn, Phys. Rev. Lett.
{\bf 38} (1977) 1440.
\bibitem{weinax} S. Weinberg, Phys. Rev. Lett. {\bf 40} (1978) 223.
\bibitem{wilcax} F. Wilczek, Phys. Rev. Lett. {\bf 40} (1978) 279.
\bibitem{DFSZ}M. Dine, W. Fischler and M. Srednicki,
    Phys. Lett. {\bf B104} (1981) 199; A.R. Zhitnitsky, Sov. J. Nucl. Phys. {\bf 31} (1980)
    260.
\bibitem{KSVZ} J.E. Kim, Phys. Rev. Lett. {\bf 43} (1979) 103;
M.A. Shifman, A.I.~Vainshtein and V.I.~Zakharov, Nucl. Phys.
{\bf B166} (1980) 493.
\bibitem{svrcek}
  P.~Svrcek and E.~Witten,
  %``Axions in string theory,''
  JHEP {\bf 0606}, 051 (2006)
  [arXiv:hep-th/0605206].
  %%CITATION = JHEPA,0606,051;%%
\bibitem{0904.0647}
%\cite{Hamann:2009yf}
%\bibitem{Hamann:2009yf}
  J.~Hamann, S.~Hannestad, G.~G.~Raffelt and Y.~Y.~Y.~Wong,
  %``Isocurvature forecast in the anthropic axion window,''
  arXiv:0904.0647 [hep-ph].
  %%CITATION = ARXIV:0904.0647;%%
\bibitem{raffelt}H. Murayama, G. Raffelt, C. Hagmann, K. van Bibber, and L.J. Rosenberg,
Eur. Phys. J. {\bf C3} (1998) 264.
\bibitem{invisible} J.E. Kim, Phys. Rev. Lett. {\bf 43} (1979) 103;
M.A. Shifman, A.I. Vainshtein and V.I. Zakharov, Nucl. Phys. {\bf B166}
(1980) 493; M. Dine, W. Fischler and M. Srednicki, Phys. Lett.
{\bf 104B} (1981) 199.
\bibitem{gondolo9}
%\cite{Visinelli:2009zm}
%\bibitem{Visinelli:2009zm}
  L.~Visinelli and P.~Gondolo,
  %``Dark Matter Axions Revisited,''
  arXiv:0903.4377 [astro-ph.CO].
  %%CITATION = ARXIV:0903.4377;%%

\bibitem{sikivie_ax} P. Sikivie, Phys. Rev. Lett.  {\bf 48} (1982) 1156.
\bibitem{bibber}
%\cite{Asztalos:2006kz}
%\bibitem{Asztalos:2006kz}
  S.~J.~Asztalos, L.~J.~Rosenberg, K.~van Bibber, P.~Sikivie and K.~Zioutas,
  %``Searches for astrophysical and cosmological axions,''
  Ann.\ Rev.\ Nucl.\ Part.\ Sci.\  {\bf 56} (2006) 293.
  %%CITATION = ARNUA,56,293;%%
\bibitem{cast} 
 K.~Zioutas {\it et al.}  [CAST Collaboration],
  %``First results from the CERN Axion Solar Telescope (CAST),''
  Phys.\ Rev.\ Lett.\  {\bf 94}, 121301 (2005)
  [arXiv:hep-ex/0411033].
  %%CITATION = PRLTA,94,121301;%%
%\bibitem{axionphoton} {\bfseries MISSING REFERENCE!!!}
\bibitem{axionphoton}
%\cite{Csaki:2001yk}
%\bibitem{Csaki:2001yk}
  C.~Csaki, N.~Kaloper and J.~Terning,
  %``Dimming supernovae without cosmic acceleration,''
  Phys.\ Rev.\ Lett.\  {\bf 88}, 161302 (2002)
  [arXiv:hep-ph/0111311];
%\cite{Ostman:2004eh}
%\bibitem{Ostman:2004eh}
  L.~\"Ostman and E.~M\"ortsell,
  %``Limiting the dimming of distant type Ia supernovae,''
  JCAP {\bf 0502}, 005 (2005)
  [arXiv:astro-ph/0410501];
  %%CITATION = JCAPA,0502,005;%%
  %%CITATION = PRLTA,88,161302;%%
%\cite{Mirizzi:2006zy}
%\bibitem{Mirizzi:2006zy}
  A.~Mirizzi, G.~G.~Raffelt and P.~D.~Serpico,
  %``Photon axion conversion in intergalactic magnetic fields and  cosmological
  %consequences,''
  Lect.\ Notes Phys.\  {\bf 741}, 115 (2008)
  [arXiv:astro-ph/0607415].
  %%CITATION = LNPHA,741,115;%%

\bibitem{murayama}
%\cite{Davoudiasl:2004be}
%\bibitem{Davoudiasl:2004be}
  H.~Davoudiasl, R.~Kitano, T.~Li and H.~Murayama,
  %``The new minimal standard model,''
  Phys.\ Lett.\  B {\bf 609}, 117 (2005)
  [arXiv:hep-ph/0405097].
  %%CITATION = PHLTA,B609,117;%%
 
\bibitem{ma}
  N.~G.~Deshpande and E.~Ma,
  %``Pattern Of Symmetry Breaking With Two Higgs Doublets,''
  Phys.\ Rev.\  D {\bf 18}, 2574 (1978); 
  %%CITATION = PHRVA,D18,2574;%%
%\cite{Deshpande:1977rw}
  E.~Ma,
  %``Verifiable radiative seesaw mechanism of neutrino mass and dark matter,''
  Phys.\ Rev.\ D {\bf 73} (2006) 077301
  [hep-ph/0601225].
  %%CITATION = HEP-PH 0601225;%%


\bibitem{rychkov}
  R.~Barbieri, L.~J.~Hall and V.~S.~Rychkov,
  Phys.\ Rev.\ D {\bf 74} (2006) 015007
  [hep-ph/0603188].
  %%CITATION = HEP-PH 0603188;%%
\bibitem{honorezlopez}
  L.~Lopez Honorez, E.~Nezri, J.~L.~Oliver and M.~H.~G.~Tytgat,
  %``The inert doublet model: An archetype for dark matter,''
  JCAP {\bf 0702}, 028 (2007)
  [arXiv:hep-ph/0612275].
  %%CITATION = JCAPA,0702,028;%%
 
\bibitem{tytgat}
      %\cite{Andreas:2009hj}
      %\bibitem{Andreas:2009hj}
        S.~Andreas, M.~H.~G.~Tytgat and Q.~Swillens,
        %``Neutrinos from Inert Doublet Dark Matter,''
        arXiv:0901.1750 [hep-ph].
        %%CITATION = ARXIV:0901.1750;%%
\bibitem{erik1}
%\cite{Gustafsson:2007pc}
  M.~Gustafsson, E.~Lundstr\"om, L.~Bergstr\"om and J.~Edsj\"o,
  %``Significant gamma lines from inert Higgs dark matter,''
  Phys.\ Rev.\ Lett.\  {\bf 99}, 041301 (2007)
  [arXiv:astro-ph/0703512].
  %%CITATION = PRLTA,99,041301;%%

\bibitem{fermi} The Fermi-LAT collaboration, see e.g. 
 %\cite{Atwood:2009ez}
%\bibitem{Atwood:2009ez}
  W.~B.~Atwood {\it et al.}  [LAT Collaboration],
  %``The Large Area Telescope on the Fermi Gamma-ray Space Telescope Mission,''
  arXiv:0902.1089 [astro-ph.IM].
  %%CITATION = ARXIV:0902.1089;%%

\bibitem{prelaunch} 
%\cite{Baltz:2008wd}
%\bibitem{Baltz:2008wd}
  E.~A.~Baltz {\it et al.},
  %``Pre-launch estimates for GLAST sensitivity to Dark Matter annihilation
  %signals,''
  JCAP {\bf 0807}, 013 (2008)
  [arXiv:0806.2911 [astro-ph]].
  %%CITATION = JCAPA,0807,013;%%

\bibitem{erik2}
      %\cite{Lundstrom:2008ai}
        E.~Lundstr\"om, M.~Gustafsson and J.~Edsj\"o,
        %``The Inert Doublet Model and LEP II Limits,''
        arXiv:0810.3924 [hep-ph].
        %%CITATION = ARXIV:0810.3924;%%
\bibitem{hut} P. Hut, Phys. Lett. {\bf 69B} (1977) 85.
\bibitem{leeweinberg} B.W. Lee and S. Weinberg, Phys. Rev. Lett.
{\bf 39} (1977) 165.
\bibitem{vysotsky} M.I. Vysotsky, A.D. Dolgov and Ya. B. Zel'dovich,
JETP Lett. {\bf 26} (1977) 188.
\bibitem{gunn} J. E. Gunn, B.W. Lee, I. Lerche, D.N. Schramm and
G. Steigman, Astrophys. J. {\bf 223} (1978) 1015.
%\cite{Wess:1974tw}
%\bibitem{Wess:1974tw}
\bibitem{elnu}
C. Amsler et al., Phys. Lett. {\bf B667}, 1 (2008).
%\cite{Hannestad:2008js}
%\bibitem{Hannestad:2008js}
\bibitem{steen} 
 S.~Hannestad, A.~Mirizzi, G.~G.~Raffelt and Y.~Y.~Y.~Wong,
  %``Cosmological constraints on neutrino plus axion hot dark matter: Update
  %after WMAP-5,''
  JCAP {\bf 0804}, 019 (2008)
  [arXiv:0803.1585 [astro-ph]].
  %%CITATION = JCAPA,0804,019;%%
\bibitem{seljak}
%\cite{Seljak:2004xh}
%\bibitem{Seljak:2004xh}
  U.~Seljak {\it et al.}  [SDSS Collaboration],
  %``Cosmological parameter analysis including SDSS Ly-alpha forest and  galaxy
  %bias: Constraints on the primordial spectrum of fluctuations,  neutrino mass,
  %and dark energy,''
  Phys.\ Rev.\  D {\bf 71}, 103515 (2005)
  [arXiv:astro-ph/0407372].
  %%CITATION = PHRVA,D71,103515;%%
\bibitem{steenh}
%\cite{Brandbyge:2008js}
%\bibitem{Brandbyge:2008js}
  J.~Brandbyge and S.~Hannestad,
  %``Grid Based Linear Neutrino Perturbations in Cosmological N-body
  %Simulations,''
  arXiv:0812.3149 [astro-ph].
  %%CITATION = ARXIV:0812.3149;%%

\bibitem{tg} S. Tremaine and J.G. Gunn, Phys. Rev. Lett. {\bf 42} (1979) 407.
\bibitem{boy}
%\cite{Boyarsky:2008ju}
%\bibitem{Boyarsky:2008ju}
%\cite{Gorbunov:2008ka}
%\bibitem{Gorbunov:2008ka}
  D.~Gorbunov, A.~Khmelnitsky and V.~Rubakov,
  %``Constraining sterile neutrino dark matter by phase-space density
  %observations,''
  JCAP {\bf 0810}, 041 (2008)
  [arXiv:0808.3910 [hep-ph]];
  %%CITATION = JCAPA,0810,041;%%
  A.~Boyarsky, O.~Ruchayskiy and D.~Iakubovskyi,
  %``A lower bound on the mass of Dark Matter particles,''
  JCAP {\bf 0903}, 005 (2009)
  [arXiv:0808.3902 [hep-ph]].
  %%CITATION = JCAPA,0903,005;%%
\bibitem{sterile} %\cite{Dodelson:1993je}
%\bibitem{Dodelson:1993je}
  S.~Dodelson and L.~M.~Widrow,
  %``Sterile Neutrinos as Dark Matter,''
  Phys.\ Rev.\ Lett.\  {\bf 72}, 17 (1994)
  [arXiv:hep-ph/9303287];
  %%CITATION = PRLTA,72,17;%%
%\cite{Shi:1998km}
%\bibitem{Shi:1998km}
  X.~D.~Shi and G.~M.~Fuller,
  %``A new dark matter candidate: Non-thermal sterile neutrinos,''
  Phys.\ Rev.\ Lett.\  {\bf 82}, 2832 (1999)
  [arXiv:astro-ph/9810076];
  %%CITATION = PRLTA,82,2832;%%
%\cite{Asaka:2006ek}
%\bibitem{Asaka:2006ek}
  T.~Asaka, M.~Shaposhnikov and A.~Kusenko,
  %``Opening a new window for warm dark matter,''
  Phys.\ Lett.\  B {\bf 638}, 401 (2006)
  [arXiv:hep-ph/0602150].
  %%CITATION = PHLTA,B638,401;%%

\bibitem{abazajian}
%\cite{Abazajian:2009hx}
%\bibitem{Abazajian:2009hx}
  K.~N.~Abazajian,
  %``Detection of Dark Matter Decay in the X-ray,''
  arXiv:0903.2040 [astro-ph.CO].
  %%CITATION = ARXIV:0903.2040;%%
\bibitem{shaposh}
%\cite{Boyarsky:2009ix}
%\bibitem{Boyarsky:2009ix}
  A.~Boyarsky, O.~Ruchayskiy and M.~Shaposhnikov,
  %``The role of sterile neutrinos in cosmology and astrophysics,''
  arXiv:0901.0011 [hep-ph].
  %%CITATION = ARXIV:0901.0011;%%
\bibitem{neulimit}
%\cite{Hansen:2001zv}
%\bibitem{Hansen:2001zv}
  S.~H.~Hansen, J.~Lesgourgues, S.~Pastor and J.~Silk,
  %``Closing the window on warm dark matter,''
  Mon.\ Not.\ Roy.\ Astron.\ Soc.\  {\bf 333}, 544 (2002)
  [arXiv:astro-ph/0106108];
%\cite{Abazajian:2006yn}
%\bibitem{Abazajian:2006yn}
  K.~Abazajian and S.~M.~Koushiappas,
  %``Constraints on sterile neutrino dark matter,''
  Phys.\ Rev.\  D {\bf 74}, 023527 (2006)
  [arXiv:astro-ph/0605271];
%\cite{Boyanovsky:2007ay}
%\bibitem{Boyanovsky:2007ay}
  D.~Boyanovsky, H.~J.~de Vega and N.~Sanchez,
  %``Constraints on dark matter particles from theory, galaxy observations and
  %N-body simulations,''
  Phys.\ Rev.\  D {\bf 77}, 043518 (2008)
  [arXiv:0710.5180 [astro-ph]];
  %%CITATION = PHRVA,D77,043518;%%
  %%CITATION = PHRVA,D74,023527;%%
%\cite{Boyarsky:2008mt}
%\bibitem{Boyarsky:2008mt}
  A.~Boyarsky, J.~Lesgourgues, O.~Ruchayskiy and M.~Viel,
  %``Realistic sterile neutrino dark matter with keV mass does not contradict
  %cosmological bounds,''
  arXiv:0812.3256 [hep-ph].
  %%CITATION = ARXIV:0812.3256;%%
\bibitem{simon}
%\cite{Simon:2007dq}
%\bibitem{Simon:2007dq}
  J.~D.~Simon and M.~Geha,
  %``The Kinematics of the Ultra-Faint Milky Way Satellites: Solving the Missing
  %Satellite Problem,''
  Astrophys.\ J.\  {\bf 670}, 313 (2007)
  [arXiv:0706.0516 [astro-ph]].
  %%CITATION = ASJOA,670,313;%%

\bibitem{strigari}
 %\cite{Strigari:2007at}
%\bibitem{Strigari:2007at}
  L.~E.~Strigari, S.~M.~Koushiappas, J.~S.~Bullock, M.~Kaplinghat, J.~D.~Simon, M.~Geha and B.~Willman,
  %``The Most Dark Matter Dominated Galaxies: Predicted Gamma-ray Signals from
  %the Faintest Milky Way Dwarfs,''
  arXiv:0709.1510 [astro-ph].
  %%CITATION = ARXIV:0709.1510;%%
\bibitem{maccio}
%\cite{Maccio':2009dx}
%\bibitem{Maccio':2009dx}
  A.~V.~Maccio', X.~Kang, F.~Fontanot, R.~S.~Somerville, S.~E.~Koposov and P.~Monaco,
  %``On the origin and properties of Ultrafaint Milky Way Satellites in a LCDM
  %Universe,''
  arXiv:0903.4681 [astro-ph.CO].
  %%CITATION = ARXIV:0903.4681;%%

\bibitem{wess}
  J.~Wess and B.~Zumino,
  %``Supergauge Transformations in Four-Dimensions,''
  Nucl.\ Phys.\  B {\bf 70}, 39 (1974).
  %%CITATION = NUPHA,B70,39;%%
\bibitem{fayet} P. Fayet, Phys. Lett. {\bf 86B} (1979) 272.
\bibitem{cabibbo} N. Cabibbo, G. Farrar and L. Maiani, Phys. Lett.
{\bf 105B} (1981) 155.
\bibitem{primack} H. Pagels and J.R. Primack, Phys. Rev. Lett.
{\bf 48} (1982) 223.
\bibitem{weinberg2} S. Weinberg, Phys. Rev. Lett. {\bf 50} (1983) 387.
\bibitem{goldberg} H. Goldberg, Phys. Rev. Lett. {\bf 50} (1983) 1419.
\bibitem{krauss} L.M. Krauss, Nucl. Phys. {\bf B227} (1983) 556.
\bibitem{ellis0} J. Ellis, J.S. Hagelin, D.V. Nanopoulos, K. Olive,
M. Srednicki, Nucl. Phys. {\bf B238} (1984) 453.
\bibitem{msugra}
%\cite{Chamseddine:1982jx}
%\bibitem{Chamseddine:1982jx}
  A.~H.~Chamseddine, R.~L.~Arnowitt and P.~Nath,
  %``Locally Supersymmetric Grand Unification,''
  Phys.\ Rev.\ Lett.\  {\bf 49}, 970 (1982);
  %%CITATION = PRLTA,49,970;%%
%\cite{Barbieri:1982eh}
%\bibitem{Barbieri:1982eh}
  R.~Barbieri, S.~Ferrara and C.~A.~Savoy,
  %``Gauge Models With Spontaneously Broken Local Supersymmetry,''
  Phys.\ Lett.\  B {\bf 119}, 343 (1982);
  %%CITATION = PHLTA,B119,343;%%
%\cite{Hall:1983iz}
%\bibitem{Hall:1983iz}
  L.~J.~Hall, J.~D.~Lykken and S.~Weinberg,
  %``Supergravity As The Messenger Of Supersymmetry Breaking,''
  Phys.\ Rev.\  D {\bf 27} (1983) 2359.
  %%CITATION = PHRVA,D27,2359;%%
\bibitem{msug_dark}
%\cite{Edsjo:2003us}
%\bibitem{Edsjo:2003us}
  J.~Edsj\"o, M.~Schelke, P.~Ullio and P.~Gondolo,
  %``Accurate relic densities with neutralino, chargino and sfermion
  %coannihilations in mSUGRA,''
  JCAP {\bf 0304}, 001 (2003)
  [arXiv:hep-ph/0301106];
  %%CITATION = JCAPA,0304,001;%%
  %\cite{Baer:2009bu}
%\bibitem{Baer:2009bu}
  H.~Baer, E.~K.~Park and X.~Tata,
  %``Collider, direct and indirect detection of supersymmetric dark matter,''
  arXiv:0903.0555 [hep-ph].
  %%CITATION = ARXIV:0903.
\bibitem{grav0}
%\cite{Takayama:2000uz}
%\bibitem{Takayama:2000uz}
  F.~Takayama and M.~Yamaguchi,
  %``Gravitino dark matter without R-parity,''
  Phys.\ Lett.\  B {\bf 485}, 388 (2000)
  [arXiv:hep-ph/0005214];
  %%CITATION = PHLTA,B485,388;%%

%\cite{Buchmuller:2008vw}
%\bibitem{Buchmuller:2008vw}
  W.~Buchmuller, M.~Endo and T.~Shindou,
  %``Superparticle Mass Window from Leptogenesis and Decaying Gravitino Dark
  %Matter,''
  JHEP {\bf 0811}, 079 (2008)
  [arXiv:0809.4667 [hep-ph]].
  %%CITATION = JHEPA,0811,079;%%


\bibitem{gravitino}


%\cite{Hamaguchi:2009sz}
%\bibitem{Hamaguchi:2009sz}
  K.~Hamaguchi, F.~Takahashi and T.~T.~Yanagida,
  %``Decaying gravitino dark matter and an upper bound on the gluino mass,''
  arXiv:0901.2168 [hep-ph];
  %%CITATION = ARXIV:0901.2168;%%


%\cite{Endo:2009by}
%\bibitem{Endo:2009by}
  M.~Endo and T.~Shindou,
  %``R-parity Violating Right-Handed Neutrino in Gravitino Dark Matter
  %Scenario,''
  arXiv:0903.1813 [hep-ph];
  %%CITATION = ARXIV:0903.1813;%%



%\cite{Chen:2009ew}
%\bibitem{Chen:2009ew}
  S.~L.~Chen, R.~N.~Mohapatra, S.~Nussinov and Y.~Zhang,
  %``R-Parity Breaking via Type II Seesaw, Decaying Gravitino Dark Matter and
  %PAMELA Positron Excess,''
  arXiv:0903.2562 [hep-ph];
  %%CITATION = ARXIV:0903.2562;%%
%\cite{Ishiwata:2009pt}
%\bibitem{Ishiwata:2009pt}
  K.~Ishiwata, S.~Matsumoto and T.~Moroi,
  %``PAMELA and ATIC Anomalies in Decaying Gravitino Dark Matter Scenario,''
  arXiv:0903.3125 [hep-ph].
  %%CITATION = ARXIV:0903.3125;%%

%\cite{Ibarra:2008jk}
%\bibitem{Ibarra:2008jk}
  A.~Ibarra and D.~Tran,
  %``Decaying Dark Matter and the PAMELA Anomaly,''
  JCAP {\bf 0902}, 021 (2009)
  [arXiv:0811.1555 [hep-ph]].
  %%CITATION = JCAPA,0902,021;%%

\bibitem{haberkane} H.E. Haber and G.L. Kane, Phys. Rep. {\bf 117} (1985) 75.
\bibitem{direct}
%\cite{Ellis:2005mb}
%\bibitem{Ellis:2005mb}
  J.~R.~Ellis, K.~A.~Olive, Y.~Santoso and V.~C.~Spanos,
  %``Update on the direct detection of supersymmetric dark matter,''
  Phys.\ Rev.\  D {\bf 71}, 095007 (2005)
  [arXiv:hep-ph/0502001].
  %%CITATION = PHRVA,D71,095007;%%

\bibitem{indirect} 
%\cite{Feng:2000zu}
%\bibitem{Feng:2000zu}
  J.~L.~Feng, K.~T.~Matchev and F.~Wilczek,
  %``Prospects for indirect detection of neutralino dark matter,''
  Phys.\ Rev.\  D {\bf 63}, 045024 (2001)
  [arXiv:astro-ph/0008115].
  %%CITATION = PHRVA,D63,045024;%%
%\cite{Jacob:1959at}
\bibitem{Jacob:1959at}
  M.~Jacob and G.~C.~Wick,
  %``On the general theory of collisions for particles with spin,''
  Annals Phys.\  {\bf 7} (1959) 404
  [Annals Phys.\  {\bf 281} (2000) 774].
  %%CITATION = APNYA,281,774;%%

%\cite{Hisano:2002fk}
\bibitem{Hisano:2002fk}
  J.~Hisano, S.~Matsumoto and M.~M.~Nojiri,
  %``Unitarity and higher-order corrections in neutralino dark matter
  %annihilation into two photons,''
  Phys.\ Rev.\  D {\bf 67}, 075014 (2003)
  [arXiv:hep-ph/0212022].
  %%CITATION = PHRVA,D67,075014;%%

%\cite{Hisano:2003ec}
\bibitem{Hisano:2003ec}
  J.~Hisano, S.~Matsumoto and M.~M.~Nojiri,
  %``Explosive dark matter annihilation,''
  Phys.\ Rev.\ Lett.\  {\bf 92}, 031303 (2004)
  [arXiv:hep-ph/0307216].
  %%CITATION = PRLTA,92,031303;%%

%\cite{Cirelli:2005uq}
\bibitem{Cirelli:2005uq}
  M.~Cirelli, N.~Fornengo and A.~Strumia,
  %``Minimal dark matter,''
  Nucl.\ Phys.\  B {\bf 753}, 178 (2006)
  [arXiv:hep-ph/0512090].
  %%CITATION = NUPHA,B753,178;%%
%\cite{Cirelli:2007xd}
\bibitem{Cirelli:2007xd}
  M.~Cirelli, A.~Strumia and M.~Tamburini,
  %``Cosmology and Astrophysics of Minimal Dark Matter,''
  Nucl.\ Phys.\  B {\bf 787}, 152 (2007)
  [arXiv:0706.4071 [hep-ph]].
  %%CITATION = NUPHA,B787,152;%%

%\cite{ArkaniHamed:2008qn}
\bibitem{ArkaniHamed:2008qn}
  N.~Arkani-Hamed, D.~P.~Finkbeiner, T.~R.~Slatyer and N.~Weiner,
  %``A Theory of Dark Matter,''
  Phys.\ Rev.\  D {\bf 79}, 015014 (2009)
  [arXiv:0810.0713 [hep-ph]].
  %%CITATION = PHRVA,D79,015014;%%

%\cite{Lattanzi:2008qa}
\bibitem{Lattanzi:2008qa}
  M.~Lattanzi and J.~I.~Silk,
  %``Can the WIMP annihilation boost factor be boosted by the Sommerfeld
  %enhancement?,''
  arXiv:0812.0360 [astro-ph].
  %%CITATION = ARXIV:0812.0360;%%
%\cite{MarchRussell:2008yu}
\bibitem{MarchRussell:2008yu}
  J.~March-Russell, S.~M.~West, D.~Cumberbatch and D.~Hooper,
  %``Heavy Dark Matter Through the Higgs Portal,''
  JHEP {\bf 0807}, 058 (2008)
  [arXiv:0801.3440 [hep-ph]].
  %%CITATION = JHEPA,0807,058;%%
%\cite{MarchRussell:2008tu}
\bibitem{MarchRussell:2008tu}
  J.~D.~March-Russell and S.~M.~West,
  %``WIMPonium and Boost Factors for Indirect Dark Matter Detection,''
  arXiv:0812.0559 [astro-ph].
  %%CITATION = ARXIV:0812.0559;%%
%\cite{Hambye:2009pw}
\bibitem{Hambye:2009pw}
  T.~Hambye, F.~S.~Ling, L.~L.~Honorez and J.~Rocher,
  %``Scalar Multiplet Dark Matter,''
  arXiv:0903.4010 [hep-ph].
  %%CITATION = ARXIV:0903.4010;%%



\bibitem{sommerfeld}
A. Sommerfeld, Annalen der Physik {\bf 403}, 257 (1931).
%\cite{Giudice:2004tc}
\bibitem{Giudice:2004tc}
  G.~F.~Giudice and A.~Romanino,
  %``Split supersymmetry,''
  Nucl.\ Phys.\  B {\bf 699}, 65 (2004)
  [Erratum-ibid.\  B {\bf 706}, 65 (2005)]
  [arXiv:hep-ph/0406088].
  %%CITATION = NUPHA,B699,65;%%
\bibitem{cirrai}
M.~Cirelli, M.~Kadastik, M.~Raidal and A.~Strumia,
  %``Model-independent implications of the e+, e-, anti-proton cosmic ray
  %spectra on properties of Dark Matter,''
  arXiv:0809.2409 [hep-ph];
  %%CITATION = ARXIV:0809.2409;%%

\bibitem{Cholis:2008wq}
  I.~Cholis, G.~Dobler, D.~P.~Finkbeiner, L.~Goodenough and N.~Weiner,
  %``The Case for a 700+ GeV WIMP: Cosmic Ray Spectra from ATIC and PAMELA,''
  arXiv:0811.3641 [astro-ph].
  %%CITATION = ARXIV:0811.3641;%%
%\cite{Borriello:2009fa}
\bibitem{Borriello:2009fa}
  E.~Borriello, A.~Cuoco and G.~Miele,
  %``Secondary radiation from the Pamela/ATIC excess and relevance for Fermi,''
  arXiv:0903.1852 [astro-ph.GA].
  %%CITATION = ARXIV:0903.1852;%%


%\cite{Bertone:2008xr}
\bibitem{Bertone:2008xr}
  G.~Bertone, M.~Cirelli, A.~Strumia and M.~Taoso,
  %``Gamma-ray and radio tests of the e+e- excess from DM annihilations,''
  arXiv:0811.3744 [astro-ph].
  %%CITATION = ARXIV:0811.3744;%%
\bibitem{bbbet}
%\cite{Bergstrom:2008ag}
%\bibitem{Bergstrom:2008ag}
  L.~Bergstr\"om, G.~Bertone, T.~Bringmann, J.~Edsj\"o and M.~Taoso,
  %``Gamma-ray and Radio Constraints of High Positron Rate Dark Matter Models
  %Annihilating into New Light Particles,''
  arXiv:0812.3895 [astro-ph].
  %%CITATION = ARXIV:0812.3895;%%
%\cite{Hooper:2007qk}

\bibitem{prokam}
%\cite{Kamionkowski:2008gj}
%\bibitem{Kamionkowski:2008gj}
  M.~Kamionkowski and S.~Profumo,
  %``Early Annihilation And Diffuse Backgrounds In Models Of Weakly Interacting
  %Massive Particles In Which The Cross Section For Pair Annihilation Is
  %Enhanced By 1/V,''
  Phys.\ Rev.\ Lett.\  {\bf 101}, 261301 (2008)
  [arXiv:0810.3233 [astro-ph]].
  %%CITATION = PRLTA,101,261301;%%

\bibitem{radiative}
%\cite{Bergstrom:1989jr}
%\bibitem{Bergstrom:1989jr}
  L.~Bergstr\"om,
  %``RADIATIVE PROCESSES IN DARK MATTER PHOTINO ANNIHILATION,''
  Phys.\ Lett.\  B {\bf 225}, 372 (1989).
  %%CITATION = PHLTA,B225,372;%%
%\cite{Baltz:2002we}
\bibitem{Baltz:2002we}
  E.~A.~Baltz and L.~Bergstr\"om,
  %``Detection of leptonic dark matter,''
  Phys.\ Rev.\  D {\bf 67}, 043516 (2003)
  [arXiv:hep-ph/0211325].
  %%CITATION = PHRVA,D67,043516;%%

\bibitem{bring}
%\cite{Bringmann:2007nk}
%\bibitem{Bringmann:2007nk}
  T.~Bringmann, L.~Bergstr\"om and J.~Edsj\"o,
  %``New Gamma-Ray Contributions to Supersymmetric Dark Matter Annihilation,''
  JHEP {\bf 0801}, 049 (2008)
  [arXiv:0710.3169 [hep-ph]].
  %%CITATION = JHEPA,0801,049;%%
\bibitem{bring2}
%\cite{Bringmann:2008kj}
%\bibitem{Bringmann:2008kj}
  T.~Bringmann, M.~Doro and M.~Fornasa,
  %``Dark Matter signals from Draco and Willman 1: Prospects for MAGIC II and
  %CTA,''
  JCAP {\bf 0901}, 016 (2009)
  [arXiv:0809.2269 [astro-ph]].
  %%CITATION = JCAPA,0901,016;%%
%\cite{Goldberg:1983nd}
\bibitem{Goldberg:1983nd}
  H.~Goldberg,
  %``Constraint on the photino mass from cosmology,''
  Phys.\ Rev.\ Lett.\  {\bf 50}, 1419 (1983).
  %%CITATION = PRLTA,50,1419;%%

\bibitem{appplus}
T.~Appelquist, H.~C.~Cheng, and B.~A.~Dobrescu,
Phys.\ Rev.\ D\ {\bf 64}, 035002 (2001), [hep-ph/0012100].
%%CITATION = HEP-PH 0012100;%%
\bibitem{matchev}
H.~C.~Cheng, K.~T.~Matchev and M.~Schmaltz,
Phys.\ Rev.\ D {\bf 66}, 036005 (2002), [hep-ph/0204342].
%%CITATION = HEP-PH 0204342;%%
\bibitem{ser}
G.~Servant and T.~M.~P.~Tait,
Nucl.\ Phys.\ B {\bf 650}, 391 (2003), [hep-ph/0206071].
%%CITATION = HEP-PH 0206071;%%
\bibitem{cheng}
H.~C.~Cheng, J.~L.~Feng and K.~T.~Matchev,
%``Kaluza-Klein dark matter,''
Phys.\ Rev.\ Lett.\  {\bf 89} (2002) 211301
[hep-ph/0207125].
\bibitem{anton}
I.~Antoniadis,
%``A Possible New Dimension At A Few Tev,''
Phys.\ Lett.\ B {\bf 246}, 377 (1990).
%%CITATION = PHLTA,B246,377;%%
%\cite{Flacke:2008ne}
\bibitem{Flacke:2008ne}
  T.~Flacke, A.~Menon and D.~J.~Phalen,
  %``Non-minimal universal extra dimensions,''
  arXiv:0811.1598 [hep-ph].
  %%CITATION = ARXIV:0811.1598;%%

\bibitem{Hooper:2007qk}
  D.~Hooper and S.~Profumo,
  %``Dark matter and collider phenomenology of universal extra dimensions,''
  Phys.\ Rept.\  {\bf 453}, 29 (2007)
  [arXiv:hep-ph/0701197].
  %%CITATION = PRPLC,453,29;%%
\bibitem{posexcess}
%\cite{Hooper:2004xn}
%\bibitem{Hooper:2004xn}
  D.~Hooper and G.~D.~Kribs,
  %``Kaluza-Klein dark matter and the positron excess,''
  Phys.\ Rev.\  D {\bf 70}, 115004 (2004)
  [arXiv:hep-ph/0406026];
  %%CITATION = PHRVA,D70,115004;%%
%\cite{Hooper:2004bq}
%\bibitem{Hooper:2004bq}
  D.~Hooper and J.~Silk,
  %``Searching for dark matter with future cosmic positron experiments,''
  Phys.\ Rev.\  D {\bf 71}, 083503 (2005)
  [arXiv:hep-ph/0409104].
  %%CITATION = PHRVA,D71,083503;%%
%\cite{Hooper:2009fj}
\bibitem{heat} 
S.W. Barwick {\it et al.}, Phys. Rev. Lett. {\bf 75}, 390-393 (1995).


% PAMELA etc %%%%%%%%%%%%%%%%%%%%%%%%%%%%

\bibitem{Hooper:2009fj}
  D.~Hooper and K.~Zurek,
  %``The PAMELA and ATIC Signals From Kaluza-Klein Dark Matter,''
  arXiv:0902.0593 [hep-ph].
  %%CITATION = ARXIV:0902.0593;%%


\bibitem{oldpos}
D. M\"{u}ller and  K. K. Tang, Astrophys. J. {\bf 312}, 183-194 (1987);
R.L. Golden {\it et al.},Astrophys. J. {\bf 457},
L103-L106 (1996);
H. Gast, J. Olzem, and S. Schael,
Proc. XLIst Rencontres de Moriond, Electroweak Interactions and
Unified Theories, 421-428 (2006);
S.W. Barwick {\it et al.}, Astrophys. J. {\bf 482}, L191-194 (1997);
J.J. Beatty, {\it et al.}, 
Phys.\ Rev.\ Lett.\  {\bf 93}, 241102-241105 (2004).
\bibitem{moskstrong}
%\cite{Moskalenko:1997gh}
%\bibitem{Moskalenko:1997gh}
  I.~V.~Moskalenko and A.~W.~Strong,
  %``Production and propagation of cosmic-ray positrons and electrons,''
  Astrophys.\ J.\  {\bf 493}, 694 (1998)
  [arXiv:astro-ph/9710124].
  %%CITATION = ASJOA,493,694;%%
\bibitem{pamela2}
%\cite{Adriani:2008zq}
%\bibitem{Adriani:2008zq}
  O.~Adriani {\it et al.},
  %``A new measurement of the antiproton-to-proton flux ratio up to 100 GeV in
  %the cosmic radiation,''
  Phys.\ Rev.\ Lett.\  {\bf 102}, 051101 (2009)
  [arXiv:0810.4994 [astro-ph]].
  %%CITATION = PRLTA,102,051101;%%

\bibitem{pulsars}
A.K. Harding and R. Ramaty, Proc. 20th ICRC, Moscow {\bf 2}, 92-95 (1987);
A. Boulares, Astrophys. J. {\bf 342}, 807 (1989);
F.A. Aharonian, A.M. Atoyan and H.J. V\"olk, Astron. Astrophys. 
{\bf 294} L41 (1995);
A.M. Atoian, F.A. Aharonian, and H.J. V\"olk, Phys. Rev. D {\bf 52}, 3265-3275 (1995);
X. Chi, K.S. Cheng, and E.C.M. Young, Astrophys. J. {\bf 459}, L83-L86 (1996).
L. Zhang, and K.S. Cheng, Astron. Astrophys. {\bf 368}, 1063-1070 (2001);
I. B\"{u}sching, O.C. de Jager, M.S. Potgieter and C. Venter,
Astrophys. J. {\bf 78}, L39-L42 (2008);

%\cite{Kawanaka:2009dk}
%\bibitem{Kawanaka:2009dk}
  N.~Kawanaka, K.~Ioka and M.~M.~Nojiri,
  %``Cosmic-Ray Electron Excess from Pulsars is Spiky or Smooth?: Continuous and
  %Multiple Electron/Positron injections,''
  arXiv:0903.3782 [astro-ph.HE];
  %%CITATION = ARXIV:0903.3782;%%
%\cite{Yuksel:2008rf}
%\bibitem{Yuksel:2008rf}
  H.~Yuksel, M.~D.~Kistler and T.~Stanev,
  %``TeV Gamma Rays from Geminga and the Origin of the GeV Positron Excess,''
  arXiv:0810.2784 [astro-ph];
  %%CITATION = ARXIV:0810.2784;%%
%\cite{Biermann:2009qi}
%\bibitem{Biermann:2009qi}
  P.~L.~Biermann, J.~K.~Becker, A.~Meli, W.~Rhode, E.~S.~Seo and T.~Stanev,
  %``Cosmic ray positrons and electrons,''
  arXiv:0903.4048 [astro-ph.HE].
  %%CITATION = ARXIV:0903.4048;%%

\bibitem{profpuls}
%\cite{Profumo:2008ms}
%\bibitem{Profumo:2008ms}
  S.~Profumo,
  %``Dissecting Pamela (and ATIC) with Occam's Razor: existing, well-known
  %Pulsars naturally account for the 'anomalous' Cosmic-Ray Electron and
  %Positron Data,''
  arXiv:0812.4457 [astro-ph].
  %%CITATION = ARXIV:0812.4457;%%

\bibitem{cholpuls}
%\cite{Malyshev:2009tw}
%\bibitem{Malyshev:2009tw}
  D.~Malyshev, I.~Cholis and J.~Gelfand,
  %``Pulsars versus Dark Matter Interpretation of ATIC/PAMELA,''
  arXiv:0903.1310 [astro-ph.HE].
  %%CITATION = ARXIV:0903.1310;%%

\bibitem{heatexp}
%\cite{Baltz:1998xv}
%\bibitem{Baltz:1998xv}
  E.~A.~Baltz and J.~Edsj\"o,
  %``Positron Propagation and Fluxes from Neutralino Annihilation in the Halo,''
  Phys.\ Rev.\  D {\bf 59}, 023511 (1999)
  [arXiv:astro-ph/9808243];
%%CITATION = PHRVA,D59,023511;%%
%\cite{Baltz:2001ir}
%\bibitem{Baltz:2001ir}
  E.~A.~Baltz, J.~Edsj\"o, K.~Freese and P.~Gondolo,
  %``The cosmic ray positron excess and neutralino dark matter,''
  Phys.\ Rev.\  D {\bf 65}, 063511 (2002)
  [arXiv:astro-ph/0109318];
  %%CITATION = PHRVA,D65,063511;%%
%\cite{Kane:2002nm}
%\bibitem{Kane:2002nm}
  G.~L.~Kane, L.~T.~Wang and T.~T.~Wang,
  %``Supersymmetry and the cosmic ray positron excess,''
  Phys.\ Lett.\  B {\bf 536}, 263 (2002)
  [arXiv:hep-ph/0202156];
  %%CITATION = PHLTA,B536,263;%%
%\bibitem{cheng}
H.~C.~Cheng, J.~L.~Feng and K.~T.~Matchev,
%``Kaluza-Klein dark matter,''
Phys.\ Rev.\ Lett.\  {\bf 89} (2002) 211301
[hep-ph/0207125];
%\cite{Brun:2007tn}
%\bibitem{Brun:2007tn}
  P.~Brun, G.~Bertone, J.~Lavalle, P.~Salati and R.~Taillet,
  %``Antiproton and Positron Signal Enhancement in Dark Matter Mini-Spikes
  %Scenarios,''
  Phys.\ Rev.\  D {\bf 76}, 083506 (2007)
  [arXiv:0704.2543 [astro-ph]].
  %%CITATION = PHRVA,D76,083506;%%

\bibitem{dm}
M.~Cirelli and A.~Strumia,
  %``Minimal Dark Matter predictions and the PAMELA positron excess,''
  arXiv:0808.3867 [astro-ph];
  %%CITATION = ARXIV:0808.3867;%%
V.~Barger, W.~Y.~Keung, D.~Marfatia and G.~Shaughnessy,
  %``PAMELA and dark matter,''
  Phys.\ Lett.\  B {\bf 672}, 141 (2009)
  [arXiv:0809.0162 [hep-ph]];
  %%CITATION = PHLTA,B672,141;%%
J.~H.~Huh, J.~E.~Kim and B.~Kyae,
  %``Two dark matter components in N_{DM}MSSM and PAMELA data,''
  arXiv:0809.2601 [hep-ph];
  %%CITATION = ARXIV:0809.2601;%%
P.~D.~Serpico,
  %``On the possible causes of a rise with energy of the cosmic ray positron
  %fraction,''
  Phys. Rev. D {\bf 79}, 021302 (2009)
  [arXiv:0810.4846 [hep-ph]];
  %%CITATION = ARXIV:0810.4846;%%
A.~E.~Nelson and C.~Spitzer,
  %``Slightly Non-Minimal Dark Matter in PAMELA and ATIC,''
  arXiv:0810.5167 [hep-ph];
  %%CITATION = ARXIV:0810.5167;%%
T.~Bringmann,
  %``Dark Matter Annihilation Signals: The Importance of Radiative
  %Corrections,''
  arXiv:0810.5304 [hep-ph];
  %%CITATION = ARXIV:0810.5304;%%
R.~Harnik and G.~D.~Kribs,
  %``An Effective Theory of Dirac Dark Matter,''
  arXiv:0810.5557 [hep-ph];
  %%CITATION = ARXIV:0810.5557;%%
D.~Feldman, Z.~Liu and P.~Nath,
  %``PAMELA Positron Excess as a Signal from the Hidden Sector,''
  arXiv:0810.5762 [hep-ph];
  %%CITATION = ARXIV:0810.5762;%%
%\cite{Hambye:2008bq}
%\bibitem{Hambye:2008bq}
  T.~Hambye,
  %``Hidden vector dark matter,''
  JHEP {\bf 0901}, 028 (2009)
  [arXiv:0811.0172 [hep-ph]];
  %%CITATION = JHEPA,0901,028;%%
Y.~Bai and Z.~Han,
  %``A Unified Dark Matter Model in sUED,''
  arXiv:0811.0387 [hep-ph];
  %%CITATION = ARXIV:0811.0387;%%
P.~J.~Fox and E.~Poppitz,
  %``Leptophilic Dark Matter,''
  arXiv:0811.0399 [hep-ph];
  %%CITATION = ARXIV:0811.0399;%%
E.~Ponton and L.~Randall,
  %``TeV Scale Singlet Dark Matter,''
  arXiv:0811.1029 [hep-ph];
  %%CITATION = ARXIV:0811.1029;%%
S.~Baek and P.~Ko,
  %``Phenomenology of $U(1)_{L_\mu - L_\tau}$ charged dark matter at PAMELA and
  %colliders,''
  arXiv:0811.1646 [hep-ph];
  %%CITATION = ARXIV:0811.1646;%%
A.~Morselli and I.~V.~Moskalenko,
  %``Status of indirect searches in the PAMELA and Fermi era,''
  arXiv:0811.3526 [astro-ph];
  %%CITATION = ARXIV:0811.3526;%%
K.~M.~Zurek,
  %``Multi-Component Dark Matter,''
  arXiv:0811.4429 [hep-ph];
  %%CITATION = ARXIV:0811.4429;%%
M.~Taoso, S.~Ando, G.~Bertone and S.~Profumo,
  %``Angular correlations in the cosmic gamma-ray background from dark matter
  %annihilation around intermediate-mass black holes,''
  arXiv:0811.4493 [astro-ph];
  %%CITATION = ARXIV:0811.4493;%%
J.~Hisano, M.~Kawasaki, K.~Kohri and K.~Nakayama,
  %``Neutrino Signals from Annihilating/Decaying Dark Matter in the Light of
  %Recent Measurements of Cosmic Ray Electron/Positron Fluxes,''
  arXiv:0812.0219 [hep-ph];
  %%CITATION = ARXIV:0812.0219;%%
E.~J.~Chun and J.~C.~Park,
  %``Dark matter and sub-GeV hidden U(1) in GMSB models,''
  arXiv:0812.0308 [hep-ph];
  %%CITATION = ARXIV:0812.0308;%%
J.~Liu, P.~f.~Yin and S.~h.~Zhu,
  %``Prospects for Detecting Neutrino Signals from Annihilating/Decaying Dark
  %Matter to Account for the PAMELA and ATIC results,''
  arXiv:0812.0964 [astro-ph];
  %%CITATION = ARXIV:0812.0964;%%
M.~Pohl,
  %``Cosmic-ray electron signatures of dark matter,''
  arXiv:0812.1174 [astro-ph];
  %%CITATION = ARXIV:0812.1174;%%
R.~Allahverdi, B.~Dutta, K.~Richardson-McDaniel and Y.~Santoso,
  %``A Supersymmetric B-L Dark Matter Model and the Observed Anomalies in the
  %Cosmic Rays,''
  arXiv:0812.2196 [hep-ph];
  %%CITATION = ARXIV:0812.2196;%%
K.~Hamaguchi, S.~Shirai and T.~T.~Yanagida,
  %``Cosmic Ray Positron and Electron Excess from Hidden-Fermion Dark Matter
  %Decays,''
  arXiv:0812.2374 [hep-ph];
  %%CITATION = ARXIV:0812.2374;%%
K.~J.~Bae, J.~H.~Huh, J.~E.~Kim, B.~Kyae and R.~D.~Viollier,
  %``White dwarf axions, PAMELA data, and flipped-SU(5),''
  arXiv:0812.3511 [hep-ph];
  %%CITATION = ARXIV:0812.3511;%%
J.~Lavalle,
  %``On the antimatter signatures of the cosmological dark matter subhalos,''
  arXiv:0812.3576 [astro-ph];
  %%CITATION = ARXIV:0812.3576;%%
P.~Grajek, G.~Kane, D.~Phalen, A.~Pierce and S.~Watson,
  %``Is the PAMELA Positron Excess Winos?,''
  arXiv:0812.4555 [hep-ph];
  %%CITATION = ARXIV:0812.4555;%%
J.~H.~Huh, J.~E.~Kim and B.~Kyae,
  %``The minimal two DM components with SUSY,''
  arXiv:0812.5004 [hep-ph];
  %%CITATION = ARXIV:0812.5004;%%
X.~J.~Bi, P.~H.~Gu, T.~Li and X.~Zhang,
  %``ATIC and PAMELA Results on Cosmic e^\pm Excesses and Neutrino Masses,''
  arXiv:0901.0176 [hep-ph];
  %%CITATION = ARXIV:0901.0176;%%
S.~C.~Park and J.~Shu,
  %``Split-UED and Dark Matter,''
  arXiv:0901.0720 [hep-ph];
  %%CITATION = ARXIV:0901.0720;%%
I.~Gogoladze, R.~Khalid, Q.~Shafi and H.~Yuksel,
  %``CMSSM Spectroscopy in light of PAMELA and ATIC,''
  arXiv:0901.0923 [hep-ph];
  %%CITATION = ARXIV:0901.0923;%%
Q.~H.~Cao, E.~Ma and G.~Shaughnessy,
  %``Dark Matter: The Leptonic Connection,''
  arXiv:0901.1334 [hep-ph];
  %%CITATION = ARXIV:0901.1334;%%
E.~Nezri, M.~H.~G.~Tytgat and G.~Vertongen,
  %``Positrons and antiprotons from inert doublet model dark matter,''
  arXiv:0901.2556 [hep-ph];
  %%CITATION = ARXIV:0901.2556;%%
J.~Mardon, Y.~Nomura, D.~Stolarski and J.~Thaler,
  %``Dark Matter Signals from Cascade Annihilations,''
  arXiv:0901.2926 [hep-ph];
  %%CITATION = ARXIV:0901.2926;%%
D.~J.~Phalen, A.~Pierce and N.~Weiner,
  %``Cosmic Ray Positrons from Annihilations into a New, Heavy Lepton,''
  arXiv:0901.3165 [hep-ph];
  %%CITATION = ARXIV:0901.3165;%%
H.-S. Goh, L. J. Hall and P. Kumar,
%``The Leptonic Higgs as a Messenger of Dark Matter.''
  arXiv:0902.0814 [hep-ph];
M. Ibe, Y. Nakayama, H. Murayama and T. T. Yanagida,
%Nambu-Goldstone Dark Matter and Cosmic Ray Electron and Positron Excess.
  arXiv:0902.2914 [hep-ph];
%\bibitem{Shirai:2009kh}
  S.~Shirai, F.~Takahashi and T.~T.~Yanagida,
  %``Decaying Hidden Gaugino as a Source of PAMELA/ATIC Anomalies,''
  arXiv:0902.4770 [hep-ph];
  %%CITATION = ARXIV:0902.4770;%
R. Allahverdi, B. Dutta, K. Richardson-McDaniel and Y. Santoso, 
%    Title: Sneutrino Dark Matter and the Observed Anomalies in Cosmic Rays
   arXiv:0902.3463 [hep-ph];
%\cite{Cheung:2009si}
%\bibitem{Cheung:2009si}
  K.~Cheung, P.~Y.~Tseng and T.~C.~Yuan,
  %``Double-action dark matter, PAMELA and ATIC,''
  arXiv:0902.4035 [hep-ph];
  %%CITATION = ARXIV:0902.4035;%%
%\cite{Roszkowski:2009sm}
%\bibitem{Roszkowski:2009sm}
  L.~Roszkowski, R.~R.~de Austri, R.~Trotta, Y.~L.~Tsai and T.~A.~Varley,
  %``Some novel features of the Non-Universal Higgs Model,''
  arXiv:0903.1279 [hep-ph];
  %%CITATION = ARXIV:0903.1279;%%
%\cite{Finkbeiner:2009mi}
%\bibitem{Finkbeiner:2009mi}
  D.~P.~Finkbeiner, T.~Slatyer, N.~Weiner and I.~Yavin,
  %``PAMELA, DAMA, INTEGRAL and Signatures of Metastable Excited WIMPs,''
  arXiv:0903.1037 [hep-ph];
  %%CITATION = ARXIV:0903.1037;%%
%\cite{Bi:2009uj}
%\bibitem{Bi:2009uj}
  X.~J.~Bi, X.~G.~He and Q.~Yuan,
  %``Parameters in a Class of Leptophilic Dark Matter Models from ATIC and
  %PAMELA,''
  arXiv:0903.0122 [hep-ph];
  %%CITATION = ARXIV:0903.0122;%%
%\cite{Ishiwata:2009vx}
%\bibitem{Ishiwata:2009vx}
  K.~Ishiwata, S.~Matsumoto and T.~Moroi,
  %``High Energy Cosmic Rays from Decaying Supersymmetric Dark Matter,''
  arXiv:0903.0242 [hep-ph];
  %%CITATION = ARXIV:0903.0242;%%
%
\bibitem{bring3}
%\cite{Bergstr\"om:2008gr}
%\bibitem{Bergstrom:2008gr}
  L.~Bergstr\"om, T.~Bringmann and J.~Edsj\"o,
  %``New Positron Spectral Features from Supersymmetric Dark Matter - a Way to
  %Explain the PAMELA Data?,''
  Phys.\ Rev.\  D {\bf 78}, 103520 (2008)
  [arXiv:0808.3725 [astro-ph]].
  %%CITATION = PHRVA,D78,103520;%%

\bibitem{hoopa}
%\cite{Hooper:2008kv}
%\bibitem{Hooper:2008kv}
  D.~Hooper, A.~Stebbins and K.~M.~Zurek,
  %``The PAMELA and ATIC Excesses From a Nearby Clump of Neutralino Dark
  %Matter,''
  arXiv:0812.3202 [hep-ph].
  %%CITATION = ARXIV:0812.3202;%%
\bibitem {lavalle}
%\cite{Bringmann:2009ip}
%\bibitem{Bringmann:2009ip}
  T.~Bringmann, J.~Lavalle and P.~Salati,
  %``Intermediate Mass Black Holes and Nearby Dark Matter Point Sources: A
  %Myth-Buster,''
  arXiv:0902.3665 [astro-ph.CO].
  %%CITATION = ARXIV:0902.3665;%%

\bibitem{thaler} 
%\cite{Nomura:2008ru}
%\bibitem{Nomura:2008ru}
  Y.~Nomura and J.~Thaler,
  %``Dark Matter through the Axion Portal,''
  arXiv:0810.5397 [hep-ph].
  %%CITATION = ARXIV:0810.5397;%%

%\cite{Hisano:2009rc}
\bibitem{Hisano:2009rc}
  J.~Hisano, M.~Kawasaki, K.~Kohri, T.~Moroi and K.~Nakayama,
  %``Cosmic Rays from Dark Matter Annihilation and Big-Bang Nucleosynthesis,''
  arXiv:0901.3582 [hep-ph].
  %%CITATION = ARXIV:0901.3582;%%
%\cite{Cholis:2008wq}


\end{thebibliography}
\end{document}